\documentclass[12pt]{article}
\usepackage[margin=1in]{geometry}
\usepackage{amsmath,amssymb,amsthm,bm}
\usepackage{graphicx}
\usepackage{booktabs,tabularx,array,multirow}
\usepackage[authoryear,round]{natbib}
\usepackage{setspace}
\usepackage[hidelinks]{hyperref}
\usepackage{xcolor}
\usepackage{float}
\usepackage[font=small,labelfont=bf]{caption}
\usepackage{enumitem}
\usepackage{adjustbox}
\usepackage{microtype}
\usepackage{xr}
\newtheorem{assumption}{Assumption}
\newtheorem{proposition}{Proposition}
\newtheorem{lemma}{Lemma}
\newtheorem{corollary}{Corollary}
\theoremstyle{definition}

\newtheorem{remark}{Remark}
\newtheorem{example}{Example}

\newcommand{\EI}{\mathrm{EI}}
\newcommand{\R}{\mathbb{R}}
\newcommand{\E}{\mathbb{E}}
\DeclareMathOperator*{\argmin}{arg\,min}
\DeclareMathOperator{\conv}{conv}
\DeclareMathOperator{\dist}{dist}
\newcommand{\RMSPE}{\mathrm{RMSPE}}

\title{\textbf{Synthetic Control under\\ Out-of-Span Common Shocks:\\ Diagnosis, Exposure Balance and Correction}}
\author{Rok Spruk\thanks{School of Economics and Business, University of Ljubljana, Kardeljeva plo\v{s}\v{c}ad 17, SI-1000 Ljubljana, Slovenia. E-mail: \texttt{rok.spruk@ef.uni-lj.si}. Replication files are available in the journal's data archive.}}
\date{\today}

\begin{document}
\maketitle
\thispagestyle{empty}

\begin{abstract}
\noindent Synthetic control relies on pre-treatment fit to balance the unobserved factors that drive untreated outcomes. The argument fails when an observed common shock, such as a commodity price, moves with the latent factors before treatment and departs from them afterwards: pre-treatment fit then says almost nothing about the synthetic unit's exposure to the shock. I derive a period-specific bound that links the post-treatment bias of any weighting estimator to pre-treatment misfit through a leverage statistic, and show that the statistic decomposes exactly into a latent-factor component and the partial leverage of the observed shock. The decomposition yields a diagnostic for such \emph{silent factors} that uses no post-treatment outcome of the treated unit. Balancing an observed exposure index removes the problem, but simplex weights can achieve balance only when the treated unit lies inside the donors' exposure hull; outside the hull, every simplex-weighted estimator carries an exposure imbalance with an explicit lower bound. For that case I propose an exposure-augmented synthetic control that corrects the remaining imbalance with a cross-donor regression, characterise what the correction identifies and when it fails, and derive bias-aware confidence intervals with a breakdown value. Monte Carlo experiments, including designs in which the correction's assumptions fail, map its gains and its limits. Iran's 2012 sanctions coincided with the 2014 oil collapse, and Iran is more oil-dependent than any of its natural comparators. The 2012--15 output loss of about 12 percent survives every adjustment. The persistent loss that standard synthetic control reports after 2016 does not: any adjustment for oil exposure removes it, and the long-run effect is not identified with these comparators.


\end{abstract}

\clearpage
\setcounter{page}{1}

\section{Introduction}\label{sec:intro}

The synthetic control method \citep{abadie2003,abadie2010,abadie2015} has become the default design for case studies in which a single country, region or firm receives an intervention. Its appeal rests on a transparent argument. If a weighted average of untreated units reproduces the treated unit's outcome path over a long pre-treatment period, it must also reproduce the treated unit's exposure to the unobserved common factors that generated that path, and so remains a valid counterfactual afterwards. \citet{abadie2010} formalise the argument with a bias bound that vanishes as the pre-treatment period lengthens, and later work has refined, relaxed and augmented it \citep{fermanpinto2021,ferman2021,benmichael2021,arkhangelsky2021,abadie2021}.

This paper studies a situation, common in applied work, in which the argument fails in a predictable way. Suppose untreated outcomes depend on an observed common shock, a commodity price say, to which units are heterogeneously exposed. Before treatment the shock moves with the latent factors: commodity prices rose through the 2000s, and so did incomes in most emerging economies. After treatment the shock departs from them. A synthetic unit can then match the treated unit's pre-treatment path almost perfectly while being far less exposed to the shock, because higher exposure and faster underlying growth are observationally equivalent before treatment. Once the shock reverses, the two diverge and the estimated treatment effect absorbs the difference. Nothing in the pre-treatment fit warns of this.

The case that motivates the paper makes the problem concrete. In early 2012 the European Union embargoed Iranian crude and Iranian banks were disconnected from SWIFT. Two years later the world oil price fell by more than half. Iran's oil rents averaged 23 percent of GDP over 1996--2011; among the nineteen developing economies that kept normal relations with the United States, Europe and Israel throughout the period, and that therefore form the natural donor pool, the most oil-dependent had 17 percent. A synthetic Iran built from these donors fits Iran's pre-2012 log income path with a root mean squared error of 0.014, yet its oil exposure is 15.5 percentage points below Iran's. Standard synthetic control estimates imply that the sanctions reduced Iran's GDP per capita by 12 percent in 2012--15 and by a persistent 8 percent thereafter. How much of the persistent part reflects sanctions, and how much Iran's greater exposure to an oil price that collapsed after 2014?

\paragraph{Contributions.} The paper makes four contributions. The ingredients are elementary; the contribution lies in assembling them into a diagnostic, an impossibility result and a correction that applied researchers can use, and in showing, on a case that matters, how much they change the answer.

First, I characterise the problem. For any weighting estimator, the post-treatment bias in period $t$ is bounded by the systematic part of the pre-treatment misfit times a leverage statistic $\kappa_t$: the Mahalanobis length of period $t$'s factor vector in the metric of the pre-treatment factor second moments (Proposition~\ref{prop:leverage}). The bound is finite-sample and period-specific, and it requires neither perfect pre-treatment fit nor a well-conditioned factor second-moment matrix, the two conditions behind the bound of \citet{abadie2010}. Over the pre-treatment period $\kappa_t^2$ averages the number of factors, so values far above that benchmark signal that the post-treatment factors have left their pre-treatment span. The same statistic governs the prediction variance of regression-based estimators; it therefore prices extrapolation for both families. When an observed shock is the source of the departure, $\kappa_t^2$ splits exactly into a latent-factor component and the squared partial leverage $\ell_t^2$ of the shock: the post-treatment residual of the shock on the latent factors, standardised by its pre-treatment root mean square (Corollary~\ref{cor:fwl}). A large $|\ell_t|$ after treatment, together with a high pre-treatment $R^2$ of the shock on the latent factors, identifies a \emph{silent factor}: one that is nearly invisible before treatment and active after it, so that pre-treatment fit leaves the synthetic unit's exposure almost unrestricted. \citet{fermanpinto2021} and \citet{ferman2021} show that imperfect fit leaves latent factors unbalanced; a silent factor is the extreme case in which even perfect fit does so, and the point of Corollary~\ref{cor:fwl} is that the shock and the donors alone suffice to detect it.

Second, I characterise the remedy and its limits. If an observed, pre-determined index such as oil rents relative to GDP captures heterogeneity in exposure, balancing the index removes the silent factor from the bias (Proposition~\ref{prop:balance}). With simplex weights, balance is feasible if and only if the treated unit lies in the convex hull of the donors' exposures. The trade-off between exposure balance and pre-treatment fit, the \emph{exposure--fit frontier}, is convex. Outside the hull, every simplex-weighted estimator has an exposure bias bounded below by the distance to the hull times the size of the out-of-span shock (Proposition~\ref{prop:geometry}). The class includes synthetic control, penalized synthetic control, synthetic difference-in-differences and de-meaned synthetic control.

Third, for the case outside the hull I propose the \emph{exposure-augmented synthetic control} (EASC). It corrects the exposure imbalance that remains after weighting with a cross-donor regression of outcome changes on exposure, controlling for donors' pre-treatment growth. It is an augmented synthetic control in the sense of \citet{benmichael2021}, with one difference that matters here: the outcome model is a regression on an index measured outside the outcome path, which is what allows it to target a factor that pre-treatment outcomes do not reveal. Proposition~\ref{prop:easc} characterises the correction. When the growth controls span the latent loadings, the coefficient identifies the out-of-span component of the shock, the part that weighting cannot handle, and the exposure term of the bias vanishes if either the weights balance exposure or the regression is unbiased. What remains is bounded by a leverage in which the small pre-treatment $\ell_{T_0}$ replaces the large post-treatment $\ell_t$. When the controls do not span the loadings, the appendix gives the extra term and the simulations measure it. The correction needs donors whose exposure brackets the treated unit's, but those donors need not receive weight. Inference uses bias-aware confidence intervals in the manner of \citet{armstrong2018,armstrong2020}: the bias bound is $M\kappa\RMSPE$, where $M$ is the share of pre-treatment misfit treated as systematic, and a breakdown value $M^*$ reports how much systematic misfit a conclusion survives (Proposition~\ref{prop:ci}), as in the relative-magnitude bounds of \citet{rambachan2023}.

Fourth, I document the method's finite-sample behaviour and apply it. Twelve Monte Carlo designs calibrated to the application cover the cases the theory distinguishes: no exposure channel; an in-span shock; out-of-span shocks with the treated unit inside and outside the donors' exposure hull; measurement error and nonlinearity in exposure; and four adverse designs built to stress the correction, with loadings correlated with exposure, a latent break that raises leverage to the level seen in the application, the treated unit at the edge of the donors' latent distribution, and a second, unobserved shock on which the treated unit loads idiosyncratically. When the shock is out of span and the treated unit lies outside the exposure hull, standard estimators are biased by three to four times the true effect. EASC removes the bias when its assumptions hold; when they fail, it removes the exposure-related part and leaves the treated-specific part, which no estimator can recover. The bias-aware intervals cover in both cases. In the application, every weighting estimator, with or without the exposure adjustment and in either donor pool, finds an output loss of about 12 percent in 2012--15, with placebo $p$-values of 0.045--0.09 and breakdown values of 0.21--0.41. The persistent loss after 2016 does not survive. Once oil exposure is accounted for, by reweighting or by regression, the 2016--17 effect is close to zero, the 2018--24 estimate changes sign, and every standard estimator that produces weights moves to the same exposure-adjusted path. The positive 2018--24 estimate is no more credible than the negative one it replaces: latent leverage after 2015 is so high that the data cannot identify the long-run effect with these comparators. What the analysis establishes is that standard synthetic control cannot be used to claim one.

\paragraph{Related literature.} The paper contributes to four literatures. The first studies the bias of synthetic control when pre-treatment fit is imperfect or the factor structure is rich \citep{abadie2010,fermanpinto2021,ferman2021,arkhangelsky2021,benmichael2021,abadielhour2021,arkhangelsky2023}. \citet{abadie2010} bound the bias under perfect fit and a factor model whose pre-treatment second-moment matrix is well conditioned. Proposition~\ref{prop:leverage} gives a finite-sample, period-specific version without either condition and shows that ill-conditioning combined with a post-treatment break is precisely the case in which the bound fails. \citet{fermanpinto2021} show that synthetic control is biased when pre-treatment fit is imperfect, and propose demeaning; \citet{ferman2021} studies the many-donor case. Their bias concerns latent factors the weights fail to match. The silent factor is the limiting case: a factor the weights cannot match because the pre-treatment data carry almost no information about it. \citet{ferman2020} show how specification choices can move synthetic-control estimates; the exposure imbalance reported here is a specification statistic that should be fixed in advance.

The second literature augments synthetic control with an outcome model \citep{benmichael2021,arkhangelsky2021,athey2021,xu2017,amjad2018,hsiao2012,doudchenko2016} or uses covariates and auxiliary outcomes \citep{abadie2010,botosaru2019,kaul2022,abadievives2022,sun2023,shi2022}. EASC is an augmented synthetic control whose outcome model is a cross-sectional regression on a pre-determined exposure index rather than on pre-treatment outcomes, which is what lets it correct imbalance in a factor that pre-treatment outcomes do not reveal. \citet{kellogg2021} trade off interpolation and extrapolation biases by combining matching with synthetic control. The exposure--fit frontier makes the same trade-off explicit in a single exposure dimension, with a lower bound on what any interpolating estimator can achieve. Regularised and penalized variants \citep{abadielhour2021,chen2026} address over-fitting and interpolation bias within the span of the pre-treatment data. The problem here lies outside that span.

The third literature concerns inference for synthetic control. It includes placebo and permutation tests \citep{abadie2010,firpo2018}, conformal inference \citep{chernozhukov2021}, prediction intervals \citep{cattaneo2021} and asymptotic approaches \citep{li2020,masini2021}. Bias-aware intervals indexed by a sensitivity parameter adapt \citet{armstrong2018,armstrong2020} and \citet{armstrong2022}, and the relative-magnitude restrictions of \citet{rambachan2023}, to synthetic control. Pre-treatment misfit plays the role of pre-trends, and leverage that of the extrapolation horizon. \citet{chernozhukov2021} propose conformal tests that are exact under exchangeability of residuals; a silent factor breaks exchangeability between pre- and post-treatment residuals, which is the case the bias-aware intervals are built for.

The fourth literature estimates the economic effects of sanctions, in particular on Iran \citep{dizaji2013,neuenkirch2015,gharehgozli2017,laudati2023,ghomi2022,farzanegan2025}. \citet{gharehgozli2017} applies synthetic control to the 2012 sanctions and finds a large and persistent output loss. \citet{laudati2023} identify sanctions shocks from newspaper coverage in a structural model of the Iranian economy and stress that oil exports are the main transmission channel. That observation is why the comparison with oil-exposed donors matters. More broadly, case-study evidence on sanctions, liberalisation, natural disasters, populism and nationalism applies synthetic control to countries that are often commodity exporters \citep{billmeier2013,cavallo2013,pinotti2015,born2019,funke2023}, and in such studies the treatment often coincides with a commodity cycle. The diagnostics proposed here can be computed for any of them at negligible cost.

\paragraph{Outline.} Section~\ref{sec:model} sets out the model and the leverage bound. Section~\ref{sec:balance} studies exposure balance and its geometry. Section~\ref{sec:estimators} introduces the estimators and inference. Section~\ref{sec:mc} reports the Monte Carlo evidence and Section~\ref{sec:app} the application. Section~\ref{sec:conclusion} concludes with practical recommendations. Proofs, additional simulations, donor documentation and further empirical results are in the Online Appendix.

\section{Model and the leverage bound}\label{sec:model}

\subsection{Setting}

There are $J+1$ units observed over $T$ periods. Unit $1$ is treated from period $T_0+1$ onwards; units $j=2,\dots,J+1$ are never treated. Let $Y_{jt}^N$ denote untreated potential outcomes and $\tau_t=Y_{1t}-Y_{1t}^N$ the treatment effect for $t>T_0$. A weighting estimator chooses $w=(w_2,\dots,w_{J+1})'$ with $\sum_j w_j=1$ and estimates
\begin{equation}\label{eq:tauhat}
\hat\tau_t(w)=Y_{1t}-\sum_{j\ge2}w_jY_{jt},\qquad t>T_0 .
\end{equation}
For $t\le T_0$ the same expression, $e_t(w)$, is the pre-treatment residual. Synthetic control restricts $w$ to the simplex $\Delta^J=\{w:w_j\ge0,\ \sum_jw_j=1\}$ and minimises $\sum_{t\le T_0}e_t(w)^2$, possibly with covariates. Penalized synthetic control, the unit weights of synthetic difference-in-differences and de-meaned synthetic control also use simplex weights, the last two with an intercept. The ridge-augmented synthetic control of \citet{benmichael2021} uses weights that sum to one but may be negative.

\begin{assumption}[Factor model with an observed common shock]\label{ass:model}
For all $j$ and $t$,
\begin{equation}\label{eq:model}
Y_{jt}^N=\delta_t+\mu_j'\lambda_t+\theta_j'x_t+\varepsilon_{jt},
\end{equation}
where $x_t\in\R^q$ is an observed common shock, $\lambda_t\in\R^r$ are latent factors whose first element is 1 (so that $\mu_j$ contains a unit fixed effect), $\theta_j\in\R^q$ is unit $j$'s exposure to the observed shock, and $\varepsilon_{jt}$ has mean zero conditional on $\{\lambda_t,x_t\}_t$, $\{\mu_j,\theta_j\}_j$ and treatment assignment.
\end{assumption}

The observed shock could be treated as one more latent factor, and many results below hold for any partition of the factors. Singling it out matters for two reasons. Its behaviour after treatment can be inspected, because it is measured. And unit-level exposure to it is often measurable as well, which opens the remedy of Section~\ref{sec:balance}. Stack $g_t=(\lambda_t',x_t')'\in\R^p$ with $p=r+q$, $\beta_j=(\mu_j',\theta_j')'$, and define the imbalance of weights $w$ and the noise term
\[
\Delta(w)=\beta_1-\sum_{j\ge2}w_j\beta_j,\qquad u_t(w)=\varepsilon_{1t}-\sum_{j\ge2}w_j\varepsilon_{jt}.
\]

\begin{lemma}[Error decomposition]\label{lem:decomp}
Under Assumption~\ref{ass:model}, for any $w$ with $\sum_jw_j=1$: $e_t(w)=\Delta(w)'g_t+u_t(w)$ for $t\le T_0$, and $\hat\tau_t(w)-\tau_t=\Delta(w)'g_t+u_t(w)$ for $t>T_0$.
\end{lemma}

The common component $\delta_t$ cancels because the weights sum to one. Pre-treatment fit is informative about the post-treatment error only through $\Delta(w)$.

\subsection{A leverage bound}

Let $S_0=T_0^{-1}\sum_{t\le T_0}g_tg_t'$ be the pre-treatment second-moment matrix of the factors, assumed nonsingular for now, and define the \emph{systematic pre-treatment misfit} and the \emph{leverage} of period $t$:
\begin{equation}\label{eq:eta-kappa}
\eta(w)=\Big(T_0^{-1}\sum_{t\le T_0}\big(\Delta(w)'g_t\big)^2\Big)^{1/2}=\big(\Delta(w)'S_0\Delta(w)\big)^{1/2},\qquad
\kappa_t=\big(g_t'S_0^{-1}g_t\big)^{1/2}.
\end{equation}
The systematic misfit is the part of the pre-treatment root mean squared prediction error (RMSPE) that does not come from noise. For fixed $w$ and independent noise, $\E[\RMSPE(w)^2]=\eta(w)^2+\sigma^2(1+\|w\|^2)$ with $\RMSPE(w)^2=T_0^{-1}\sum_{t\le T_0}e_t(w)^2$, so in expectation the RMSPE overstates $\eta(w)$. For weights fitted to the pre-treatment period this is no longer guaranteed; Section~\ref{sec:inference} returns to the point. For a set $\mathcal T$ of post-treatment periods write $\bar g_{\mathcal T}=|\mathcal T|^{-1}\sum_{t\in\mathcal T}g_t$ and $\kappa_{\mathcal T}=(\bar g_{\mathcal T}'S_0^{-1}\bar g_{\mathcal T})^{1/2}$.

\begin{proposition}[Leverage bound]\label{prop:leverage}
Under Assumption~\ref{ass:model}, for every $w$ with $\sum_jw_j=1$ and every $t>T_0$,
\begin{equation}\label{eq:bound}
|\Delta(w)'g_t|\le\kappa_t\,\eta(w),
\end{equation}
and $\Delta(w)'g_t=\E[\hat\tau_t(w)-\tau_t\mid\beta,g]$ is the conditional bias for any $w$ that is fixed or independent of the post-treatment noise. The same bound holds for averages over $\mathcal T$ with $\kappa_{\mathcal T}\le|\mathcal T|^{-1}\sum_{t\in\mathcal T}\kappa_t$ in place of $\kappa_t$. The bound is attained: for every $t$ and $\eta>0$ there is an imbalance vector $\Delta$ with $\Delta'S_0\Delta=\eta^2$ and $|\Delta'g_t|=\kappa_t\eta$ (whether some $w\in\Delta^J$ produces that $\Delta$ depends on the donors). Moreover, $T_0^{-1}\sum_{t\le T_0}\kappa_t^2=p$, and $\kappa_t^2/T_0$ is the prediction-variance factor of period $t$ (its out-of-sample leverage) in a least-squares regression of pre-treatment outcomes on $g_t$.
\end{proposition}

The inequality is algebraic and holds for estimated weights as well; reading it as a bias requires only that the weights not depend on post-treatment noise, which holds for weights estimated from pre-treatment data when the errors are serially independent. Proposition~\ref{prop:leverage} turns the intuition that synthetic control ``works if the post-treatment period looks like the pre-treatment period'' into a number. A given amount of systematic misfit becomes post-treatment bias at rate $\kappa_t$. Before treatment, $\kappa_t^2$ averages $p$ by construction, and in a regression with $T_0$ observations a point with $\kappa_t^2/T_0$ above $2p/T_0$ or $3p/T_0$ is conventionally called high-leverage \citep{belsley1980}; the same thresholds, $\kappa_t>\sqrt{2p}$ or $\sqrt{3p}$, serve here. In the application $\kappa_t$ exceeds $\sqrt{3p}\approx3.5$ from 2012 and reaches 31 by 2024. The last statement of the proposition connects the bound to regression-based estimators. If the treated unit's pre-treatment outcomes are regressed on the factors, as in interactive fixed effects \citep{xu2017,gobillon2016} or the panel-data approach of \citet{hsiao2012}, the prediction variance in period $t$ is $\sigma^2\kappa_t^2/T_0$. Weighting estimators pay for extrapolation in bias of order $\kappa_t\eta$; regression estimators pay in variance of order $\kappa_t^2/T_0$. Both prices rise with leverage.

The proof is a Cauchy--Schwarz inequality in the $S_0$ metric; \citet{benmichael2021} use the same device to bound the bias of their augmented estimator by imbalance times factor magnitude. What is new is the object to which it is applied and what follows from that. The bound of \citet{abadie2010} assumes perfect pre-treatment fit and a smallest eigenvalue of $S_0$ bounded away from zero, and vanishes as $T_0\to\infty$. Here pre-treatment fit need not be perfect, $S_0$ may be nearly singular, and the bound is specific to the post-treatment period. Near-singularity of $S_0$ is not a technicality. It is the defining feature of the situation this paper studies, and Corollary~\ref{cor:fwl} shows that it can be diagnosed.

\subsection{Silent factors}

Partition the factors into the latent block $\lambda_t$ and the observed shock $x_t$, and project the shock on the latent factors over the pre-treatment period:
\begin{equation}\label{eq:proj}
x_t=A'\lambda_t+r_t,\qquad A=\Big(\sum_{t\le T_0}\lambda_t\lambda_t'\Big)^{-1}\sum_{t\le T_0}\lambda_tx_t',
\end{equation}
which defines the residual $r_t$ for \emph{all} $t$, with $\sum_{t\le T_0}\lambda_tr_t'=0$. Let $S_\lambda=T_0^{-1}\sum_{t\le T_0}\lambda_t\lambda_t'$, $S_r=T_0^{-1}\sum_{t\le T_0}r_tr_t'$, $\kappa^\lambda_t=(\lambda_t'S_\lambda^{-1}\lambda_t)^{1/2}$, and define the \emph{partial leverage} of the shock $\ell_t=S_r^{-1/2}r_t$. Write
\begin{equation}\label{eq:composite}
\tilde\mu_j=\mu_j+A\theta_j
\end{equation}
for unit $j$'s \emph{composite} loading on the latent factors, and $\tilde\Delta(w)=\tilde\mu_1-\sum_jw_j\tilde\mu_j$ and $\Delta_\theta(w)=\theta_1-\sum_jw_j\theta_j$ for the imbalances in composite loadings and in exposure.

\begin{corollary}[Silent factors]\label{cor:fwl}
Under Assumption~\ref{ass:model}, model \eqref{eq:model} can be written as $Y_{jt}^N=\delta_t+\tilde\mu_j'\lambda_t+\theta_j'r_t+\varepsilon_{jt}$, and:
\begin{enumerate}[label=(\alph*),itemsep=0pt]
\item $\kappa_t^2=(\kappa^\lambda_t)^2+\|\ell_t\|^2$ for every $t$;
\item for every $w$, $\Delta(w)'g_t=\tilde\Delta(w)'\lambda_t+\Delta_\theta(w)'r_t$ and $\eta(w)^2=\tilde\Delta(w)'S_\lambda\tilde\Delta(w)+\Delta_\theta(w)'S_r\Delta_\theta(w)$;
\item consequently, for $q=1$, any weights with $\eta(w)\le\bar\eta$ satisfy $|\Delta_\theta(w)|\le\bar\eta/S_r^{1/2}$. Weights with $\tilde\Delta(w)=0$ have systematic misfit $\eta(w)=|\Delta_\theta(w)|S_r^{1/2}$ and post-treatment bias $\Delta_\theta(w)r_t$, so that $|\text{bias}_t|/\eta(w)=|\ell_t|$.
\end{enumerate}
\end{corollary}

The reparametrisation in Corollary~\ref{cor:fwl} is the heart of the problem. The part of the shock that moves with the latent factors before treatment, $A'\lambda_t$, is absorbed into the composite loadings, which pre-treatment fit does balance. The remainder $r_t$ is a \emph{silent factor}. It contributes only $S_r$, the variance of the shock that the latent factors do not explain, to pre-treatment outcomes, yet it can be large after treatment. When the pre-treatment $R^2$ of $x_t$ on $\lambda_t$ is close to one, $S_r$ is small, and part (c) shows that pre-treatment fit then barely restricts the exposure imbalance $\Delta_\theta(w)$. Matching composite loadings suffices to fit the pre-treatment path; a synthetic unit with lower exposure and faster underlying growth fits as well as one with the right exposure. After treatment, bias accrues at rate $|\ell_t|$ per unit of systematic misfit.

\begin{example}[Two donors]\label{ex:two}
Let $\lambda_t=(1,f_t)'$ with $f_t$ a linear trend, and let $x_t=\varrho f_t$ exactly before treatment, so that $A=(0,\varrho)'$, $S_r=0$ and $S_0$ is singular (the limiting case of Corollary~\ref{cor:fwl}). Donor A has exposure $\theta_A=\theta_1$ and loadings $\mu_A=\mu_1$, an exact twin of the treated unit. Donor B has $\theta_B=0$ and $\mu_B=\mu_1+(0,\varrho\theta_1)'$. Both have composite loading $\tilde\mu=\mu_1+(0,\varrho\theta_1)'$, so both reproduce the treated unit's pre-treatment path exactly, and any convex combination fits perfectly. After treatment, a synthetic control that puts weight $w_B$ on B has bias $w_B\theta_1(x_t-\varrho f_t)=w_B\theta_1r_t$. The bias is zero for the twin and $\theta_1r_t$ for the unexposed donor; pre-treatment fit cannot tell them apart. Exposure balance selects the twin.
\end{example}

Both components of Corollary~\ref{cor:fwl}(a) can be computed from data. $\ell_t$ requires only the observed shock and an estimate of the latent factors, which the donors, untreated throughout, supply (Section~\ref{sec:diagnostics}). The verdict of $\ell_t$ is insensitive to the number of factors retained, whereas $\kappa^\lambda_t$ grows with it; Section~\ref{sec:app-inference} reports the consequences. A pre-treatment $R^2$ close to one together with post-treatment $|\ell_t|$ well above 2 is the signature of a silent factor. The diagnostic depends only on the shock and the donors, not on the treated unit's post-treatment outcomes, so it can be computed before the results are seen.

\section{Exposure balance}\label{sec:balance}

\subsection{Balance removes the silent factor}

Corollary~\ref{cor:fwl} shows that the dangerous component of the bias is $\Delta_\theta(w)'r_t$. Pre-treatment outcomes cannot reveal $\Delta_\theta(w)$, so information about exposure must come from elsewhere. In many applications it does: oil rents relative to GDP measure exposure to oil prices, trade shares exposure to partner demand, and sectoral employment shares exposure to industry shocks.

\begin{assumption}[Exposure index]\label{ass:index}
$\theta_j=\Gamma s_j$ for a known, pre-determined index $s_j\in\R^m$ and an unknown $q\times m$ matrix $\Gamma$.
\end{assumption}

Under Assumption~\ref{ass:index}, $\Delta_\theta(w)=\Gamma\,\EI(w)$, where
\begin{equation}\label{eq:EI}
\EI(w)=s_1-\sum_{j\ge2}w_js_j
\end{equation}
is the \emph{exposure imbalance} of $w$. It is observable for any estimator that produces weights.

\begin{proposition}[Exposure balance]\label{prop:balance}
Under Assumptions~\ref{ass:model} and~\ref{ass:index}, if $\EI(w)=0$ then for $t>T_0$
\[
\hat\tau_t(w)-\tau_t=\tilde\Delta(w)'\lambda_t+u_t(w)\qquad\text{and}\qquad|\tilde\Delta(w)'\lambda_t|\le\kappa^\lambda_t\,\eta(w).
\]
\end{proposition}

Balance removes the silent factor from the bias, however large the post-treatment shock, and replaces $\kappa_t$ by the latent leverage $\kappa^\lambda_t\le\kappa_t$. The price is that balance constrains the weights, which in general worsens pre-treatment fit and raises $\eta(w)$.

\subsection{Geometry: feasibility, the frontier and a lower bound}

\begin{proposition}[Feasibility, frontier and impossibility]\label{prop:geometry}
Let $\mathcal S=\{s_2,\dots,s_{J+1}\}$ and $d_{\min}=\min_{w\in\Delta^J}\|\EI(w)\|=\dist(s_1,\conv\mathcal S)$.
\begin{enumerate}[label=(\alph*),itemsep=0pt]
\item Exact balance with simplex weights is feasible if and only if $s_1\in\conv\mathcal S$. For scalar exposure, $d_{\min}=\max\{0,\,s_1-\max_js_j,\,\min_js_j-s_1\}$.
\item The exposure--fit frontier $R(d)=\min\{\RMSPE(w):w\in\Delta^J,\ \|\EI(w)\|\le d\}$, $d\ge d_{\min}$, is non-increasing and convex.
\item Under Assumptions~\ref{ass:model}--\ref{ass:index} with $q=m=1$ and $\theta_j=\gamma s_j$, every $w\in\Delta^J$ satisfies $\hat\tau_t(w)-\tau_t=\tilde\Delta(w)'\lambda_t+\gamma\,\EI(w)\,r_t+u_t(w)$ with $|\EI(w)|\ge d_{\min}$. In particular, if $\tilde\Delta(w)'\lambda_t=0$, the absolute bias is at least $|\gamma r_t|\,d_{\min}$. For estimators with a unit-level intercept (de-meaned SC, synthetic DiD), the same holds with $r_t$ replaced by its deviation from the (weighted) pre-treatment mean.
\end{enumerate}
\end{proposition}

Part (a) is elementary but has a sharp implication for donor-pool design. Comparators that match the treated unit's political or institutional characteristics, as sanctions studies require, are typically less exposed to the relevant commodity than the treated unit. Part (b) makes the trade-off between balance and fit well behaved: moving along the frontier towards balance raises the RMSPE at an increasing rate. Because $R(d)^2$ is then also convex and $|\EI|=d$ wherever the constraint binds, the selection criterion \eqref{eq:tuning} below is convex in $d$ and its minimiser is easy to find. Part (c) is an impossibility result. Outside the hull, no estimator that interpolates between donors can avoid an exposure bias of order $|\gamma r_t|d_{\min}$ unless an imbalance in the latent loadings happens to offset it; nothing in the estimation targets such an offset, so it cannot be relied upon. Estimators that do not restrict weights to the simplex can achieve balance, but only by extrapolating: the augmented synthetic control through negative weights, and generalized synthetic control, matrix completion and robust synthetic control through estimated loadings. Corollary~\ref{cor:fwl} shows that this extrapolation is weakly identified precisely when the silent factor matters.

\subsection{Correcting the remaining imbalance}

When balance is infeasible or too costly in fit, an outcome model can correct the remaining exposure imbalance, as in augmented synthetic control \citep{benmichael2021}. The outcome model here is a cross-donor regression on the exposure index, not a regression on pre-treatment outcomes. For a base period $b\le T_0$ (in practice $b=T_0$) and each $t$, let $\phi_t$ be the coefficient on $s_j$ in the population regression across donors of
\begin{equation}\label{eq:phi-reg}
Y_{jt}-Y_{jb}\quad\text{on}\quad(1,\ s_j',\ c_j'),
\end{equation}
where $c_j\in\R^k$ collects pre-treatment summaries of donor $j$'s own outcomes, measured before $b$. In the application they are the donor's growth within each third of the pre-treatment years before $b$. The \emph{exposure-augmented} estimator for weights $w$ is
\begin{equation}\label{eq:easc}
\hat\tau^{\,\mathrm{EA}}_t(w)=Y_{1t}-\sum_{j\ge2}w_jY_{jt}-\hat\phi_t'\,\EI(w),
\end{equation}
where $\hat\phi_t$ is the least-squares estimate of $\phi_t$.

\begin{assumption}[Spanning controls]\label{ass:span}
Let $\lambda^{-}_t$ denote the non-constant latent factors and $\tilde\mu_j^{-}$ the composite loadings on them. Across donors, $\E[\tilde\mu_j^{-}\mid s_j,c_j]=\alpha_0+\Pi c_j$ and $\E[\varepsilon_{jt}-\varepsilon_{jb}\mid s_j,c_j]=0$, and donor $j$'s loading deviation $\tilde\mu_j^{-}-\E[\tilde\mu_j^{-}\mid s_j,c_j]$ and noise difference are mean-independent of the other donors' $(s_k,c_k)$.
\end{assumption}

Assumption~\ref{ass:span} says that donors' pre-treatment growth over sub-periods, which reflects their composite latent loadings, is a sufficient control for those loadings once exposure is held fixed. The fixed-effect loading drops out of outcome changes, so only the non-constant factors matter. The assumption holds exactly in the noiseless factor model when the shock is exactly in span before treatment ($S_r=0$) and the sub-period changes of the non-constant latent factors have full rank. When $S_r>0$ it holds only approximately: the controls also load on exposure through the pre-treatment movements of $r_t$, and the coefficient in part (a) below acquires an additional term proportional to those movements (Online Appendix, Remark~\ref{A-rem:approx}). The term is small precisely when the shock is silent before treatment. The controls end before the base period so that they share no noise with the dependent variable; with serially correlated errors the separation is approximate, and the simulations, which use AR(1) errors, assess the consequences.

\begin{proposition}[Exposure-augmented synthetic control]\label{prop:easc}
Under Assumptions~\ref{ass:model}--\ref{ass:span}:
\begin{enumerate}[label=(\alph*),itemsep=0pt]
\item the regression coefficient identifies the out-of-span component of the shock: $\phi_t=\Gamma'(r_t-r_b)$;
\item for any $w$ with $\sum_jw_j=1$ and $t>T_0$,
\begin{equation}\label{eq:easc-error}
\hat\tau^{\,\mathrm{EA}}_t(w)-\tau_t=\underbrace{\tilde\Delta(w)'\lambda_t+\EI(w)'\Gamma'r_b}_{\text{latent and base-period}}\ +\ \underbrace{\EI(w)'(\phi_t-\hat\phi_t)}_{\text{correction error}}\ +\ u_t(w);
\end{equation}
\item (double robustness) for fixed $w$, the out-of-span exposure term relative to the base period, $\EI(w)'\Gamma'(r_t-r_b)$, becomes $\EI(w)'(\phi_t-\E[\hat\phi_t\mid\mathcal C])$ after augmentation, where $\mathcal C=\{s_j,c_j\}_{j\ge2}$. It is zero if $\EI(w)=0$, whatever the exposure regression, or if $\hat\phi_t$ is conditionally unbiased, which holds under Assumption~\ref{ass:span};
\item the first term in \eqref{eq:easc-error} satisfies $|\tilde\Delta(w)'\lambda_t+\EI(w)'\Gamma'r_b|\le\eta(w)\,\kappa^{\mathrm{EA}}_t$ with $\kappa^{\mathrm{EA}}_t=\big((\kappa^\lambda_t)^2+\|\ell_b\|^2\big)^{1/2}$.
\end{enumerate}
\end{proposition}

Part (a) is the key to the construction. Synthetic control has already absorbed the in-span part of the shock, $A'\lambda_t$, through the composite loadings, so the correction must target only the out-of-span part $r_t$. Conditioning on pre-treatment growth does exactly this. Without the controls, the coefficient on exposure equals $\Gamma'(r_t-r_b)$ plus the projection of the composite loadings on exposure times $\lambda_t-\lambda_b$; if, for instance, the structural loadings $\mu_j$ are unrelated to exposure, the unconditional coefficient recovers the total change $\Gamma'(x_t-x_b)$ and over-corrects. Part (d) shows what is gained. The worst-case bias of the exposure-augmented estimator is governed by $\kappa^{\mathrm{EA}}_t$, in which the base-period value $\ell_b$, a pre-treatment quantity of order one, replaces the post-treatment partial leverage $\ell_t$.

\begin{remark}[Support for the correction, not for the weights]\label{rem:support}
Proposition~\ref{prop:easc} relies on linearity of exposure (Assumption~\ref{ass:index}). If exposure is nonlinear, $\theta_j=\theta(s_j)$, and $s_1$ lies outside the range of the donors' $s_j$, then $\hat\phi_t'\EI(w)$ extrapolates the linear fit, and the error in \eqref{eq:easc-error} acquires a term of the form $[\theta(s_1)-\sum_jw_j\theta(s_j)-\gamma^{\text{lin}}\EI(w)]\,r_t$, where $\gamma^{\text{lin}}$ is the slope of the best linear approximation to $\theta(\cdot)$ over the donors' exposures. The term typically grows with the distance between $s_1$ and the support of the regression. The remedy is to include in regression \eqref{eq:phi-reg} donors whose exposure brackets $s_1$, even if they are poor matches and receive no weight. In the application, Azerbaijan and Angola play this role: they receive zero synthetic-control weight but place Iran inside the support of the exposure regression.
\end{remark}

\subsection{Relation to existing approaches}\label{sec:relation}

Four alternatives suggest themselves.

\emph{Covariates.} \citet{abadie2010} match on covariates as well as pre-treatment outcomes, with predictor weights chosen to minimise pre-treatment MSPE. Exposure can be entered as a covariate, but two problems remain. Covariate matching is still simplex weighting, so it cannot balance exposure outside the hull (Proposition~\ref{prop:geometry}(c)). And when the shock is silent before treatment, pre-treatment MSPE gives the data-driven predictor weights no reason to weight exposure; Section~\ref{sec:app-diag} shows this in the application. \citet{botosaru2019} and \citet{kaul2022} show related problems with covariates in synthetic control.

\emph{Outcome-model augmentation.} The ridge-augmented synthetic control of \citet{benmichael2021} corrects imbalance in pre-treatment outcomes. A silent factor leaves almost no trace in pre-treatment outcomes, so this correction cannot target it. EASC uses the same augmentation logic with a different outcome model: a cross-section regression on an index measured outside the outcome path.

\emph{Interpolation and extrapolation.} \citet{kellogg2021} combine nearest-neighbour matching with synthetic control to trade interpolation bias against extrapolation bias. The exposure--fit frontier makes this trade-off explicit along one dimension, exposure. Proposition~\ref{prop:geometry}(c) gives the lower bound on what interpolation can achieve when the treated unit lies outside the hull.

\emph{Residualising.} One could estimate $\gamma$ from a two-way fixed-effects regression of donor outcomes on $s_jx_t$ and apply synthetic control to the residualised outcomes. With $\gamma$ known this would work. But before treatment $s_jx_t$ is nearly collinear with $s_jA'\lambda_t$, so a regression that does not control for the latent factors confounds $\gamma$ with the correlation between exposure and latent loadings, and one that does control for them identifies $\gamma$ only from the out-of-span variation. The controlled cross-sectional regression \eqref{eq:phi-reg} uses exactly that variation, period by period, and targets only the component that weighting misses (Proposition~\ref{prop:easc}(a)).

\section{Estimators, diagnostics and inference}\label{sec:estimators}

\subsection{Diagnostics}\label{sec:diagnostics}

The latent factors are estimated from the donors, who are untreated in every period. Let $\tilde Y$ be the $T\times J$ matrix of donor outcomes, demeaned by unit and by period. To separate the latent factors from the exposure channel, I remove in each period the cross-sectional projection of $\tilde Y_{t\cdot}$ on standardised exposure and take the first $k$ left singular vectors of the result, scaled by their singular values, as $\hat\lambda_t$ (with a constant appended). This gives $\hat\kappa^\lambda_t$, $\hat\ell_t$ (from the regression of $x_t$ on $\hat\lambda_t$ over $t\le T_0$) and $\hat\kappa_t=((\hat\kappa^\lambda_t)^2+\hat\ell_t^2)^{1/2}$. The diagnostics reported in the application are:
\begin{enumerate}[itemsep=0pt]
\item the pre-treatment $R^2$ of the shock on the latent factors and the path of $\hat\ell_t$;
\item the leverage of each post-treatment phase, $\hat\kappa_{\mathcal T}$, for synthetic control and $\hat\kappa^{\mathrm{EA}}_{\mathcal T}$ for the exposure-augmented estimator;
\item the exposure imbalance $\EI(\hat w)$ of every estimator that produces weights, and the distance to the hull $d_{\min}$;
\item the exposure--fit frontier $R(d)$.
\end{enumerate}
None uses the treated unit's post-treatment outcomes.

\subsection{Estimators}

The paper compares three members of one family, two of which carry the results. All three use simplex weights $w(d)$ on the exposure--fit frontier,
\begin{equation}\label{eq:frontier}
w(d)=\argmin_{w\in\Delta^J}\sum_{t\le T_0}e_t(w)^2\quad\text{s.t.}\quad\|\EI(w)\|\le d,
\end{equation}
a convex quadratic programme. Setting $d$ at or above the synthetic-control imbalance returns the synthetic-control weights $\hat w$.
\begin{itemize}[itemsep=1pt]
\item \textbf{EBSC-min}, the exposure-balanced synthetic control at $d=d_{\min}$: the best-fitting weights among those with the smallest attainable imbalance. Inside the hull it balances exactly; outside, it does as well as simplex weights can.
\item \textbf{EASC}, the exposure-augmented synthetic control \eqref{eq:easc} at the synthetic-control weights $\hat w$.
\item \textbf{EASC$^*$}, the exposure-augmented estimator at $w(\hat d)$, where $\hat d$ minimises an estimate of post-treatment mean squared error along the frontier:
\begin{equation}\label{eq:tuning}
\hat d=\argmin_{d}\ \frac1{T-T_0}\sum_{t>T_0}\Big[(\hat\kappa^{\lambda}_t)^2\big(R(d)^2-R(\bar d)^2\big)+\widehat{\mathrm{se}}(\hat\phi_t)^2\,\EI(w(d))^2\Big],
\end{equation}
where $\bar d=\|\EI(\hat w)\|$. The criterion is a heuristic estimate of post-treatment mean squared error, included to show how the frontier can be used; in the application it selects the synthetic-control end of the frontier, so EASC$^*$ coincides with EASC. The first term treats the extra pre-treatment misfit that balance induces, $R(d)^2-R(\bar d)^2$, as systematic and scales it by the latent leverage, following Proposition~\ref{prop:easc}(d). The second is the variance of the correction. Balancing pays when the correction is imprecise and the cost in fit is small.
\end{itemize}
$\hat\phi_t$ is estimated by least squares in \eqref{eq:phi-reg} with $b=T_0$, and its standard error by the delete-one-donor jackknife. EASC can be applied to the weights of any estimator that produces them; Table~\ref{tab:estimators} below uses this to put the standard estimators on a common, exposure-adjusted footing.

\subsection{Inference}\label{sec:inference}

Inference concerns averages of $\tau_t$ over pre-specified post-treatment phases $\mathcal T$. Write the estimation error of an estimator as bias plus noise, $\bar\tau-\tau_{\mathcal T}=\mathsf{b}_{\mathcal T}+v_{\mathcal T}$. By Propositions~\ref{prop:leverage} and~\ref{prop:easc}, $|\mathsf{b}_{\mathcal T}|\le\kappa_{\mathcal T}\eta(w)$ for synthetic control and $|\mathsf{b}_{\mathcal T}|\le\kappa^{\mathrm{EA}}_{\mathcal T}\eta(w)$ for EASC. The systematic misfit $\eta(w)$ is not separately identified from noise. The RMSPE is the natural observable proxy, so I index the bias bound by
\begin{equation}\label{eq:B}
B_{\mathcal T}(M)=M\,\kappa_{\mathcal T}\,\RMSPE(\hat w),\qquad M\in[0,1],
\end{equation}
where $M$ is the share of the pre-treatment RMSPE treated as systematic and aligned with the post-treatment factor direction. $M=0$ gives the usual interval that ignores bias. $M=1$ treats the whole pre-treatment misfit as systematic and adversarial, which by Proposition~\ref{prop:leverage} is the worst case when $\eta(w)\le\RMSPE(w)$. For estimated weights, $\RMSPE(\hat w)$ is an optimistic proxy for $\eta(\hat w)$, particularly when $J$ is large relative to $T_0$, so values of $M$ above 1 cannot be excluded. $M$ is best read as a reporting scale.

\begin{proposition}[Bias-aware intervals and breakdown]\label{prop:ci}
Suppose $v_{\mathcal T}\sim N(0,\sigma^2)$ and $|\mathsf{b}_{\mathcal T}|\le B$. Let $\mathrm{cv}_\alpha(\rho)$ be the $1-\alpha$ quantile of $|Z+\rho|$, $Z\sim N(0,1)$. Then $\mathrm{CI}(B)=[\bar\tau\pm\sigma\,\mathrm{cv}_\alpha(B/\sigma)]$ satisfies $\Pr(\tau_{\mathcal T}\in\mathrm{CI}(B))\ge1-\alpha$, with equality when $|\mathsf{b}_{\mathcal T}|=B$. The half-length is increasing in $B$, so the breakdown value $M^*=\sup\{M\ge0:0\notin\mathrm{CI}(B_{\mathcal T}(M))\}$ is well defined and solves $|\bar\tau|=\sigma\,\mathrm{cv}_\alpha(M^*\kappa_{\mathcal T}\RMSPE/\sigma)$ when $0\notin\mathrm{CI}(0)$.
\end{proposition}

The interval follows \citet{armstrong2018,armstrong2020} and is shorter than adding the bias bound to a conventional interval. Its nominal level is conditional on the bias bound: the interval at a given $M$ covers at $1-\alpha$ if the systematic share of the misfit is at most $M$, and the exercise is to report how the conclusion changes as $M$ moves. Because $\eta(\hat w)$ is not identified, no single $M$ delivers an unconditional 95 percent interval; the breakdown value $M^*$, not any one interval, is the summary statistic. The breakdown value parallels the relative-magnitude bounds of \citet{rambachan2023}. In difference-in-differences, post-treatment violations are bounded by a multiple of pre-treatment violations; here post-treatment bias is bounded by a multiple of pre-treatment misfit, scaled by leverage. $M^*$ answers a concrete question: what share of the pre-treatment misfit would have to be systematic, and aligned with the post-treatment factor direction, to overturn the conclusion?

The noise standard deviation $\sigma$ is estimated from in-space placebos. Each donor in turn is treated as the treated unit, with the remaining donors as controls. Each placebo's phase average is divided by its own pre-treatment RMSPE, and the standard deviation of these ratios is rescaled by the treated unit's RMSPE. This follows the post/pre ratio logic of \citet{abadie2010} and prevents poorly fitted placebos from dominating. A robust version replaces the standard deviation by $1.4826$ times the median absolute deviation. For EASC, $|\EI(\hat w)|\cdot\overline{\mathrm{se}}(\hat\phi_{\mathcal T})$ is added in quadrature to account for the correction error; this ignores the dependence between the placebo gaps and $\hat\phi_t$, which share donors, and the simulations check whether it matters. The noise scale is estimated from $J$ placebos and $\sigma$ is treated as known in Proposition~\ref{prop:ci}, so with $J$ around 20 the intervals are approximate. I also report the conventional placebo $p$-value based on $|\bar\tau|/\RMSPE$, whose smallest attainable value is $1/(J+1)$. When a silent factor is active, the placebo distribution is not a pure noise distribution, and nothing in it reveals the treated unit's exposure bias. The simulations below show that intervals which ignore bias then under-cover badly for synthetic control, and that the bias-aware intervals restore coverage.

\section{Monte Carlo evidence}\label{sec:mc}


\subsection{Design}\label{sec:mc-design}

The simulations reproduce the main features of the application, with one deliberate exception: the baseline designs keep latent leverage moderate so that the exposure channel can be isolated, and design S9 then raises it to the level seen in the application. There are $T=29$ years (1996--2024), and treatment starts in 2012. The observed shock $x_t$ is the actual log Brent price, centred on its 1996--2011 mean. Untreated outcomes follow
\begin{equation}\label{eq:dgp}
Y_{jt}^N=a_j+d_t+\theta_jx_t+\mu_jf_t+\nu_jh_t+\varepsilon_{jt},
\end{equation}
with a latent trend $f_t=(t-2003.5)/10$ and a latent random walk $h_t$ with innovation s.d.\ 0.03, redrawn in each replication. The errors $\varepsilon_{jt}$ are AR(1) with coefficient 0.5 and marginal s.d.\ 0.015; $d_t=0.02(t-1996)$ and $a_j\sim N(0,0.3^2)$ is a unit effect. Exposure is linear, $\theta_j=\gamma s_j$ with $\gamma=0.008$, and $s_j$ is the actual 1996--2011 mean oil rent of each donor. The trend loading is $\mu_j=\xi_j-\gamma\varrho s_j$ with $\xi_j\sim N(0,0.08^2)$, where $\varrho=1.32$ is the pre-2012 slope of $x_t$ on $f_t$. Composite loadings $\tilde\mu_j=\xi_j$ are therefore unrelated to exposure, and pre-treatment fit carries no information about it: the situation of Corollary~\ref{cor:fwl}. The loadings on $h_t$ are $\nu_j\sim N(0,1)$. The treated unit has $s_1=22.87$, Iran's value, and $a_1=\xi_1=\nu_1=0$, so it lies at the centre of the donors' latent loadings. The treatment effect is $-0.10$ in 2012--15 and $-0.05$ thereafter.

The calibration matches the application on the dimensions that matter. The pre-2012 $R^2$ of $x_t$ on the latent trend is 0.89 (0.92 in the application). The average partial leverage of $x_t$ is $-$6.7 in 2016--17 and $-$7.8 in 2018--24 (in the application, $\hat\ell_t$ averages $-$7.7 and $-$8.7). The implied out-of-span exposure effect $\phi_t=\gamma(r_t-r_{2011})$ averages $-$0.0123 and $-$0.0142 per percentage point of oil rents in the two phases, against estimates of $-$0.0049 and $-$0.0128 in the application.

There are twelve designs:
\begin{description}[itemsep=0pt,font=\normalfont\itshape]
\item[S0] no exposure channel ($\gamma=0$);
\item[S1] in-span shock: after 2012, $x_t$ follows its pre-2012 relation with $f_t$;
\item[S2] baseline: the actual Brent path, with the treated unit outside the hull of the 19 donor exposures (maximum 16.56);
\item[S3] as S2 with 21 donors, adding exposures of 26.52 and 38.42, so that the treated unit is inside the hull;
\item[S4] as S2, with every unit's exposure observed with multiplicative log-normal error (s.d.\ 0.3);
\item[S5] as S2, with concave exposure $\theta_j=\gamma c(1-e^{-s_j/c})$, $c=20$, while the estimators assume linearity;
\item[S6] as S5 with the 21-donor pool;
\item[S7] as S2 with 38 donors (each exposure value twice, with independent loadings);
\item[S8] as S3, but composite trend loadings are correlated with exposure, $\xi_j=N(0,0.08^2)+0.004(s_j-\bar s)$, so that a regression of outcome changes on exposure alone would be confounded;
\item[S9] as S3, but the latent random walk's innovation standard deviation rises from 0.03 to 0.30 after 2014, which produces post-2015 latent leverage of the order seen in the application;
\item[S10] S8 and S9 combined, with the treated unit's trend loading two standard deviations above the donor mean, so that it lies at the edge of the donors' latent distribution as well;
\item[S11] as S3, with a second, unobserved out-of-span shock $z_t$ (zero before 2015, falling to $-1$ by 2024) whose loadings are $0.004s_j$ plus noise for donors, while the treated unit's loading is 0.06 above what its exposure predicts: a treated-specific component that no estimator can recover.
\end{description}
Designs S0--S3 and S7 satisfy the assumptions of the correction, S4--S6 bend the exposure index, and S8--S11 are built to stress the latent structure. Each design has 500 replications. The estimators are synthetic control, ridge-augmented SC, synthetic DiD, de-meaned SC, generalized SC, matrix completion, EBSC-min, EASC and EASC$^*$, tuned as in the application. For synthetic control and EASC, the full inference procedure of Section~\ref{sec:inference}, including all in-space placebos, is run in every replication.

\subsection{Bias and precision}

\begin{figure}[htbp]\centering
\includegraphics[width=\textwidth]{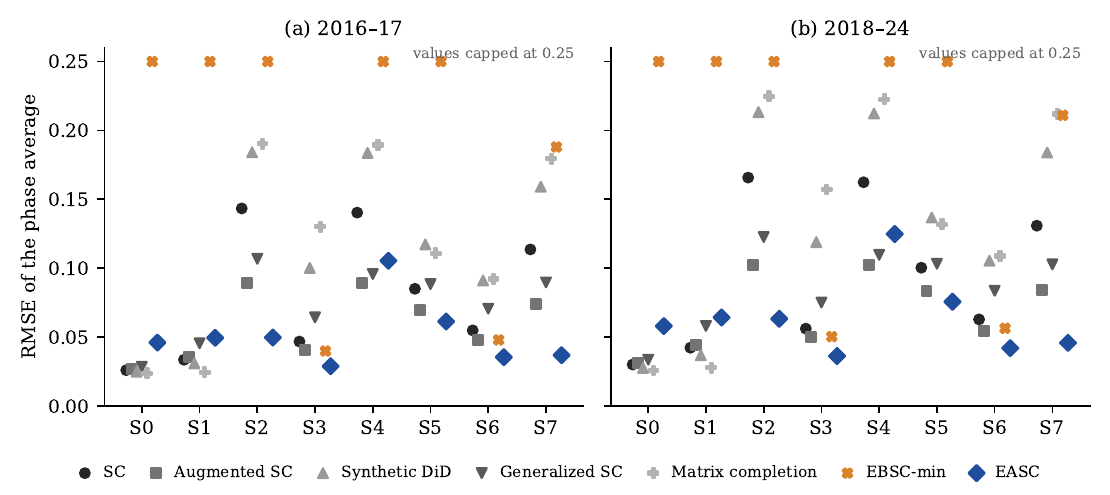}
\caption{Monte Carlo: RMSE of the 2016--17 and 2018--24 averages by design. Values above 0.25 are plotted at 0.25.}
\label{fig:mc-rmse}
\end{figure}

Table~\ref{tab:mc-bias} reports bias and RMSE of the 2018--24 average; Figure~\ref{fig:mc-rmse} shows RMSE for both later phases and Table~\ref{A-tab:mc-bias-app} the earlier ones. Four results stand out.

First, without a silent factor (S0, S1), every estimator is approximately unbiased. EASC then pays for estimating a correction it does not need: its RMSE is 0.058 in S0 and 0.064 in S1, against 0.030 and 0.042 for synthetic control. The diagnostics distinguish S1 from the out-of-span designs: the post-treatment maximum of $|\hat\ell_t|$ averages 1.9 in S1 and 9.7 in S2 (Table~\ref{A-tab:mc-diag}). In S0 the shock is out of span but outcomes do not depend on it. The exposure regression then estimates $\phi_t\approx0$, and EASC differs from SC only by noise.

Second, when the shock is out of span and the treated unit lies outside the exposure hull (S2), every standard estimator that weights donors is badly biased. The 2018--24 bias is $-$0.159 for synthetic control, $-$0.210 for synthetic DiD, $-$0.140 for de-meaned SC and $-$0.223 for matrix completion. These biases are three to four and a half times the true effect of $-0.05$, and all carry the sign that the application's synthetic control implies. Augmented SC ($-$0.082) and generalized SC ($-$0.055) partly offset the bias by extrapolating. EBSC-min fails as Proposition~\ref{prop:geometry} predicts: balancing as far as the simplex allows destroys the fit (RMSE 0.369). EASC is unbiased ($-$0.003), with an RMSE of 0.063, well under half that of synthetic control (0.166).

Third, when the treated unit is inside the hull (S3), the problem is milder, because synthetic control can draw on exposed donors (its average imbalance falls from 11.7 to 2.5). Exposure balance now works: EBSC-min has bias $-$0.004 and RMSE 0.050, and EASC and EASC$^*$ have the smallest RMSE of all estimators (0.036). With twice as many donors (S7), EASC's RMSE falls to 0.046, against 0.131 for synthetic control, as the exposure regression becomes more precise.

Fourth, EASC degrades gracefully when its assumptions are strained. Measurement error in exposure (S4) leaves its bias small ($-$0.011) but raises its RMSE to 0.125, still below synthetic control's 0.162. Under concave exposure with the treated unit outside the hull (S5), the linear correction extrapolates beyond the donors' support and over-corrects (bias 0.035). When the support covers the treated unit (S6), the bias falls to $-$0.020 and the RMSE to 0.042, against 0.063 for synthetic control. This is Remark~\ref{rem:support} at work: what matters is support for the correction, not for the weights.

Table~\ref{tab:mc-adverse} reports the four adverse designs next to S3, their common baseline. Correlation between composite loadings and exposure (S8) does no harm: the growth controls recover the loadings whatever their correlation with exposure, as Assumption~\ref{ass:span} requires, and EASC's 2018--24 bias is 0.001 with RMSE 0.034, against $-$0.030 and 0.052 for synthetic control. The design shows why the controls belong in the regression. When latent leverage is high (S9), the mean 2018--24 $\hat\kappa$ for synthetic control is 26.5, against 10.1 in S3, and every estimator's RMSE rises. EASC remains close to unbiased ($-$0.003), but its RMSE (0.152) no longer falls below that of synthetic control (0.140): the variance of the correction adds to a noise level that now dominates. The combined design (S10) is the hardest: EASC has bias 0.074 and RMSE 0.250, synthetic control 0.036 and 0.245, EBSC-min 0.108 and 0.305. Here the treated unit sits at the edge of the donors' latent distribution while the exposure imbalance remains large, and no estimator does well. EASC's bias carries the sign of the correction it applies, the pattern to expect when the weights cannot match the treated unit's loadings and the correction is scaled by a large imbalance. The hidden-shock design (S11) separates what the correction can and cannot do. The exposure regression absorbs the exposure-related part of the unobserved shock without needing to know what the shock is; the treated-specific part, worth $-0.042$ in the 2018--24 average, remains in every estimator. EASC's bias is $-$0.046, essentially that irreducible part, against $-$0.083 for synthetic control and $-$0.236 for matrix completion.

\begin{table}[htbp]\centering\small
\caption{Monte Carlo: bias and RMSE of the 2018--24 average effect}\label{tab:mc-bias}
\setlength{\tabcolsep}{4pt}\begin{adjustbox}{max width=\textwidth}\begin{tabular}{@{}lcccccccc@{}}\toprule
 & S0 & S1 & S2 & S3 & S4 & S5 & S6 & S7\\
 & \scriptsize No exposure & \scriptsize In-span shock & \scriptsize Outside hull & \scriptsize Inside hull & \scriptsize Noisy exposure & \scriptsize Concave, outside & \scriptsize Concave, inside & \scriptsize 38 donors\\\midrule
\multicolumn{9}{@{}l}{\emph{Bias, 2018--24 average}}\\
\quad SC & 0.000 & 0.003 & $-$0.159 & $-$0.034 & $-$0.154 & $-$0.091 & $-$0.050 & $-$0.124\\
\quad Augmented SC & $-$0.001 & 0.003 & $-$0.082 & $-$0.018 & $-$0.081 & $-$0.067 & $-$0.036 & $-$0.063\\
\quad Synthetic DiD & 0.001 & 0.001 & $-$0.210 & $-$0.113 & $-$0.208 & $-$0.133 & $-$0.101 & $-$0.180\\
\quad De-meaned SC & 0.000 & 0.003 & $-$0.140 & $-$0.025 & $-$0.136 & $-$0.081 & $-$0.043 & $-$0.113\\
\quad Generalized SC & $-$0.001 & 0.004 & $-$0.055 & $-$0.016 & $-$0.043 & $-$0.063 & $-$0.042 & $-$0.031\\
\quad Matrix completion & 0.001 & 0.003 & $-$0.223 & $-$0.154 & $-$0.220 & $-$0.129 & $-$0.106 & $-$0.210\\
\quad EBSC-min & $-$0.002 & $-$0.006 & $-$0.086 & $-$0.004 & $-$0.097 & $-$0.017 & $-$0.025 & $-$0.079\\
\quad EASC & 0.000 & $-$0.004 & $-$0.003 & $-$0.002 & $-$0.011 & 0.035 & $-$0.020 & 0.001\\
\quad EASC$^*$ & 0.000 & $-$0.003 & $-$0.003 & $-$0.002 & $-$0.014 & 0.034 & $-$0.022 & 0.001\\
\multicolumn{9}{@{}l}{\emph{RMSE, 2018--24 average}}\\
\quad SC & 0.030 & 0.042 & 0.166 & 0.056 & 0.162 & 0.100 & 0.063 & 0.131\\
\quad Augmented SC & 0.031 & 0.044 & 0.102 & 0.050 & \textbf{0.103} & 0.083 & 0.054 & 0.084\\
\quad Synthetic DiD & 0.028 & 0.037 & 0.213 & 0.119 & 0.212 & 0.137 & 0.105 & 0.184\\
\quad De-meaned SC & 0.030 & 0.044 & 0.148 & 0.051 & 0.144 & 0.092 & 0.058 & 0.121\\
\quad Generalized SC & 0.034 & 0.058 & 0.123 & 0.075 & 0.110 & 0.103 & 0.084 & 0.103\\
\quad Matrix completion & \textbf{0.026} & \textbf{0.028} & 0.225 & 0.157 & 0.223 & 0.132 & 0.109 & 0.212\\
\quad EBSC-min & 0.346 & 0.358 & 0.369 & 0.050 & 0.336 & 0.358 & 0.056 & 0.211\\
\quad EASC & 0.058 & 0.064 & \textbf{0.063} & 0.036 & 0.125 & 0.076 & 0.042 & 0.046\\
\quad EASC$^*$ & 0.050 & 0.064 & 0.064 & \textbf{0.036} & 0.125 & \textbf{0.074} & \textbf{0.042} & \textbf{0.046}\\
\midrule
\quad Mean $\kappa$ (SC), 2018--24 & 9.8 & 6.0 & 9.9 & 10.1 & 9.8 & 10.1 & 9.8 & 10.0\\
\bottomrule\end{tabular}\end{adjustbox}
\par\smallskip\begin{minipage}{0.97\textwidth}\footnotesize\emph{Notes.} 500 replications per design. Errors are phase averages of $\hat\tau_t-\tau_t$; the true effect is $-0.10$ in 2012--15 and $-0.05$ afterwards. Bold: smallest RMSE in the column. Designs: S0 no exposure channel ($\gamma=0$); S1 post-2012 oil price on its pre-2012 relation with the latent trend; S2 actual Brent, treated exposure 22.87 against donor exposures of the 19 normalisers (maximum 16.56); S3 as S2 with 21 donors (adds 26.52 and 38.42); S4 as S2 with exposure observed with multiplicative log-normal error (s.d.\ 0.3); S5 as S2 with concave exposure $\theta_j=\gamma c(1-e^{-s_j/c})$, $c=20$; S6 as S5 with 21 donors; S7 as S2 with 38 donors; S8--S11 are the adverse designs of Section~\ref{sec:mc-design}.\end{minipage}
\end{table}

\begin{table}[htbp]\centering\small
\caption{Monte Carlo: adverse designs that stress the exposure correction}\label{tab:mc-adverse}
\setlength{\tabcolsep}{4pt}\begin{adjustbox}{max width=\textwidth}\begin{tabular}{@{}lccccc@{}}\toprule
 & S3 & S8 & S9 & S10 & S11\\
 & \scriptsize Inside hull & \scriptsize Loadings $\sim$ exposure & \scriptsize Latent break & \scriptsize S8+S9, treated out & \scriptsize Hidden shock\\\midrule
\multicolumn{6}{@{}l}{\emph{Bias, 2012--15 average}}\\
\quad SC & $-$0.010 & $-$0.009 & $-$0.010 & 0.019 & $-$0.012\\
\quad Augmented SC & $-$0.005 & $-$0.006 & $-$0.006 & 0.003 & $-$0.008\\
\quad Synthetic DiD & $-$0.023 & $-$0.018 & $-$0.028 & 0.015 & $-$0.024\\
\quad De-meaned SC & $-$0.007 & $-$0.007 & $-$0.006 & 0.017 & $-$0.010\\
\quad Generalized SC & $-$0.003 & $-$0.003 & $-$0.008 & $-$0.002 & $-$0.007\\
\quad Matrix completion & $-$0.047 & $-$0.026 & $-$0.045 & 0.016 & $-$0.052\\
\quad EBSC-min & $-$0.002 & 0.000 & 0.001 & 0.044 & $-$0.002\\
\quad EASC & 0.001 & 0.002 & 0.000 & 0.031 & $-$0.002\\
\quad EASC$^*$ & 0.000 & 0.002 & 0.000 & 0.031 & $-$0.002\\
\multicolumn{6}{@{}l}{\emph{RMSE, 2012--15 average}}\\
\quad SC & 0.023 & 0.022 & 0.027 & 0.047 & 0.024\\
\quad Augmented SC & 0.022 & 0.023 & 0.026 & \textbf{0.029} & 0.022\\
\quad Synthetic DiD & 0.032 & 0.029 & 0.040 & 0.049 & 0.033\\
\quad De-meaned SC & 0.022 & 0.021 & \textbf{0.025} & 0.041 & 0.023\\
\quad Generalized SC & 0.028 & 0.030 & 0.036 & 0.053 & 0.029\\
\quad Matrix completion & 0.051 & 0.031 & 0.051 & 0.031 & 0.055\\
\quad EBSC-min & 0.030 & 0.031 & 0.037 & 0.075 & 0.030\\
\quad EASC & 0.021 & 0.021 & 0.026 & 0.052 & 0.020\\
\quad EASC$^*$ & \textbf{0.021} & \textbf{0.021} & 0.026 & 0.052 & \textbf{0.020}\\
\multicolumn{6}{@{}l}{\emph{Bias, 2018--24 average}}\\
\quad SC & $-$0.034 & $-$0.030 & $-$0.033 & 0.036 & $-$0.083\\
\quad Augmented SC & $-$0.018 & $-$0.018 & $-$0.017 & 0.014 & $-$0.067\\
\quad Synthetic DiD & $-$0.113 & $-$0.091 & $-$0.115 & 0.022 & $-$0.179\\
\quad De-meaned SC & $-$0.025 & $-$0.023 & $-$0.019 & 0.040 & $-$0.072\\
\quad Generalized SC & $-$0.016 & $-$0.015 & $-$0.016 & $-$0.028 & $-$0.068\\
\quad Matrix completion & $-$0.154 & $-$0.097 & $-$0.177 & 0.021 & $-$0.236\\
\quad EBSC-min & $-$0.004 & 0.000 & 0.001 & 0.108 & $-$0.043\\
\quad EASC & $-$0.002 & 0.001 & $-$0.003 & 0.074 & $-$0.046\\
\quad EASC$^*$ & $-$0.002 & 0.001 & $-$0.002 & 0.074 & $-$0.046\\
\multicolumn{6}{@{}l}{\emph{RMSE, 2018--24 average}}\\
\quad SC & 0.056 & 0.052 & 0.140 & 0.245 & 0.098\\
\quad Augmented SC & 0.050 & 0.048 & 0.139 & 0.157 & 0.085\\
\quad Synthetic DiD & 0.119 & 0.097 & 0.169 & 0.277 & 0.183\\
\quad De-meaned SC & 0.051 & 0.047 & \textbf{0.128} & 0.228 & 0.088\\
\quad Generalized SC & 0.075 & 0.081 & 0.212 & 0.268 & 0.114\\
\quad Matrix completion & 0.157 & 0.101 & 0.237 & \textbf{0.149} & 0.238\\
\quad EBSC-min & 0.050 & 0.045 & 0.191 & 0.305 & 0.065\\
\quad EASC & 0.036 & 0.034 & 0.152 & 0.250 & 0.060\\
\quad EASC$^*$ & \textbf{0.036} & \textbf{0.034} & 0.149 & 0.249 & \textbf{0.060}\\
\midrule
\quad Mean $\kappa$ (SC), 2018--24 & 10.1 & 10.0 & 26.5 & 25.4 & 10.2\\
\bottomrule\end{tabular}\end{adjustbox}
\par\smallskip\begin{minipage}{0.97\textwidth}\footnotesize\emph{Notes.} 500 replications per design; true effects $-0.10$ (2012--15) and $-0.05$ (2018--24). S3 is the common baseline (21 donors, treated unit inside the exposure hull). S8: composite trend loadings correlated with exposure, $\xi_j=N(0,0.08^2)+0.004(s_j-\bar s)$. S9: innovation s.d.\ of the latent random walk rises from 0.03 to 0.30 after 2014. S10: S8 and S9 with the treated unit's trend loading two s.d.\ above the donor mean. S11: a second, unobserved shock, zero before 2015 and $-1$ by 2024, with donor loadings $0.004s_j+N(0,0.03^2)$ and a treated loading 0.06 above that. Bold: smallest RMSE in the column.\end{minipage}
\end{table}


\subsection{Inference}

\begin{figure}[htbp]\centering
\includegraphics[width=\textwidth]{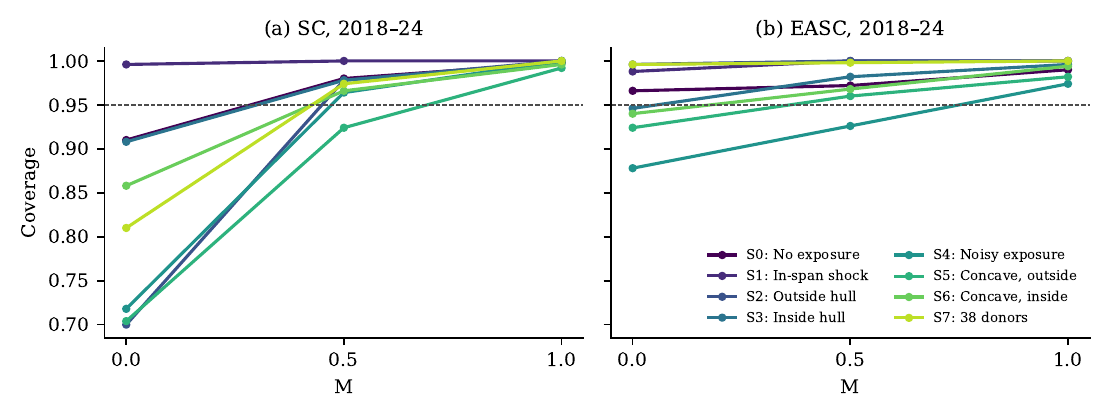}
\caption{Monte Carlo: coverage of 95\% bias-aware intervals for the 2018--24 average as a function of $M$. Dashed line: nominal level.}
\label{fig:mc-coverage}
\end{figure}

Table~\ref{tab:mc-cover} and Figure~\ref{fig:mc-coverage} report the coverage of the bias-aware intervals. For synthetic control, intervals that ignore bias ($M=0$) under-cover whenever the silent factor is active: coverage of the 2018--24 effect is 0.70 in S2, 0.72 in S4 and 0.70 in S5. Even in S0, where only latent drift matters, it is 0.91. Coverage rises with $M$, to 0.90--1.00 at $M=0.5$ and 0.98--1.00 at $M=1$ across designs, as Proposition~\ref{prop:leverage} implies. The price is length: in S2 the average interval for 2018--24 is 0.56 log points long at $M=0.5$ and 0.78 at $M=1$. EASC intervals at $M=0$ already cover in most designs (0.94--1.00 across S0--S3, S6 and S7), because the correction removes the silent-factor bias and its estimation error enters the variance. They are also shorter than the synthetic-control intervals that achieve coverage (in S2, 0.35 at $M=0$ and 0.43 at $M=0.5$, against 0.56 for synthetic control at $M=0.5$ in S2). Under measurement error in exposure (S4), EASC coverage at $M=0$ falls to 0.88 and rises to 0.93 at $M=0.5$ and 0.97 at $M=1$. The robust noise scale gives similar results (Table~\ref{A-tab:mc-cover-mad}). In the adverse designs (Table~\ref{A-tab:mc-cover-adverse}), EASC coverage of the 2018--24 effect at $M=0$ is 0.95, 0.97, 0.97 and 0.80 in S8--S11; at $M=0.5$ it is 0.97, 0.98, 1.00 and 0.90. Synthetic control at $M=0$ covers 0.94, 0.94, 0.98 and 0.75. In S8--S10 the intervals do what they are meant to do: when the assumptions behind an estimator fail, the bias-aware interval at a moderate $M$ still covers, at the cost of length. In S11 the treated-specific loading is invisible to the pre-treatment fit, so no multiple of the pre-treatment misfit bounds it, and coverage at $M=0.5$ stays below nominal for both estimators. That is the case the diagnostics cannot detect, and the reason the application treats its later-phase estimates as unidentified rather than corrected.

\begin{table}[htbp]\centering\small
\caption{Monte Carlo: coverage and length of bias-aware confidence intervals}\label{tab:mc-cover}
\setlength{\tabcolsep}{4pt}\begin{adjustbox}{max width=\textwidth}\begin{tabular}{@{}llcccccccc@{}}\toprule
 & & S0 & S1 & S2 & S3 & S4 & S5 & S6 & S7\\\midrule
\multicolumn{10}{@{}l}{\emph{Coverage of 95\% bias-aware intervals, 2016--17}}\\
\quad SC & $M=0$ & 0.87 & 0.99 & 0.65 & 0.90 & 0.68 & 0.69 & 0.84 & 0.81\\
\quad  & $M=0.5$ & 0.97 & 1.00 & 0.96 & 0.98 & 0.96 & 0.92 & 0.96 & 0.97\\
\quad  & $M=1$ & 1.00 & 1.00 & 1.00 & 1.00 & 1.00 & 0.99 & 0.99 & 1.00\\
\quad EASC & $M=0$ & 0.94 & 0.99 & 0.99 & 0.92 & 0.85 & 0.90 & 0.91 & 1.00\\
\quad  & $M=0.5$ & 0.96 & 1.00 & 1.00 & 0.97 & 0.91 & 0.94 & 0.95 & 1.00\\
\quad  & $M=1$ & 0.98 & 1.00 & 1.00 & 1.00 & 0.96 & 0.97 & 0.99 & 1.00\\
\multicolumn{10}{@{}l}{\emph{Average length, 2016--17}}\\
\quad SC & $M=0$ & 0.092 & 0.228 & 0.326 & 0.175 & 0.328 & 0.203 & 0.169 & 0.304\\
\quad  & $M=0.5$ & 0.149 & 0.293 & 0.462 & 0.243 & 0.462 & 0.292 & 0.236 & 0.412\\
\quad  & $M=1$ & 0.220 & 0.393 & 0.650 & 0.339 & 0.649 & 0.414 & 0.330 & 0.568\\
\quad EASC & $M=0$ & 0.197 & 0.272 & 0.275 & 0.125 & 0.349 & 0.212 & 0.142 & 0.234\\
\quad  & $M=0.5$ & 0.211 & 0.329 & 0.333 & 0.157 & 0.403 & 0.246 & 0.172 & 0.281\\
\quad  & $M=1$ & 0.242 & 0.425 & 0.430 & 0.207 & 0.500 & 0.307 & 0.222 & 0.362\\
\multicolumn{10}{@{}l}{\emph{Coverage of 95\% bias-aware intervals, 2018--24}}\\
\quad SC & $M=0$ & 0.91 & 1.00 & 0.70 & 0.91 & 0.72 & 0.70 & 0.86 & 0.81\\
\quad  & $M=0.5$ & 0.98 & 1.00 & 0.98 & 0.98 & 0.96 & 0.92 & 0.97 & 0.97\\
\quad  & $M=1$ & 1.00 & 1.00 & 1.00 & 1.00 & 1.00 & 0.99 & 1.00 & 1.00\\
\quad EASC & $M=0$ & 0.97 & 0.99 & 1.00 & 0.95 & 0.88 & 0.92 & 0.94 & 1.00\\
\quad  & $M=0.5$ & 0.97 & 1.00 & 1.00 & 0.98 & 0.93 & 0.96 & 0.97 & 1.00\\
\quad  & $M=1$ & 0.99 & 1.00 & 1.00 & 1.00 & 0.97 & 0.98 & 0.99 & 1.00\\
\multicolumn{10}{@{}l}{\emph{Average length, 2018--24}}\\
\quad SC & $M=0$ & 0.115 & 0.287 & 0.393 & 0.213 & 0.393 & 0.251 & 0.204 & 0.361\\
\quad  & $M=0.5$ & 0.182 & 0.378 & 0.557 & 0.296 & 0.555 & 0.361 & 0.287 & 0.493\\
\quad  & $M=1$ & 0.268 & 0.514 & 0.785 & 0.414 & 0.780 & 0.511 & 0.402 & 0.683\\
\quad EASC & $M=0$ & 0.261 & 0.348 & 0.353 & 0.160 & 0.428 & 0.283 & 0.179 & 0.289\\
\quad  & $M=0.5$ & 0.279 & 0.427 & 0.432 & 0.205 & 0.503 & 0.330 & 0.221 & 0.356\\
\quad  & $M=1$ & 0.320 & 0.558 & 0.564 & 0.274 & 0.634 & 0.415 & 0.291 & 0.466\\
\bottomrule\end{tabular}\end{adjustbox}
\par\smallskip\begin{minipage}{0.92\textwidth}\footnotesize\emph{Notes.} Intervals of Proposition~\ref{prop:ci} with bias bound $M\hat\kappa\,\RMSPE$ and noise s.d.\ from standardised in-space placebos (Section~\ref{sec:inference}); for EASC the jackknife variance of the correction is added. Nominal coverage 0.95. Designs as in Table~\ref{tab:mc-bias}.\end{minipage}
\end{table}


In sum, the simulations support four practical conclusions. Pre-treatment fit and placebo inference give no warning when a silent factor is present; the partial-leverage diagnostic does. Exposure augmentation removes the resulting bias at a moderate cost in variance, provided the exposure regression covers the treated unit. Bias-aware intervals are needed for synthetic control, and advisable for EASC, whenever leverage is high. And when the treated unit is unusual in ways that neither exposure nor pre-treatment growth reveal, EASC trades a large known bias for a smaller unknown one: an improvement, not a solution.


\section{Application: the output cost of Iran's 2012 sanctions}\label{sec:app}

\subsection{Treatment, outcome and donors}

Sanctions on Iran tightened in steps from 2006, when the UN Security Council imposed the first measures aimed at its nuclear programme. The step that cut Iran off from its main export market and from the international payments system came in early 2012. The EU adopted an embargo on Iranian crude in January 2012 (in force from July), SWIFT disconnected designated Iranian banks in March, and US legislation of December 2011 targeted foreign financial institutions dealing with the Central Bank of Iran. I set $T_0+1=2012$, the first year in which financial disconnection and the oil embargo operated together. Earlier measures and domestic events, including the 2009 political crisis and the December 2010 subsidy reform, belong to the pre-treatment period; the in-time placebos in Section~\ref{sec:app-robust} examine them. The post-treatment period runs to 2024 and is divided into three phases fixed in advance: the shock of 2012--15, the relief of 2016--17 under the Joint Comprehensive Plan of Action, and the siege of 2018--24 after the US withdrawal and the end of waivers for buyers of Iranian oil. Table~\ref{A-tab:timeline} in the Online Appendix dates each measure.

The outcome is log GDP per capita at purchasing-power parity in constant 2021 dollars (World Development Indicators), 1996--2024. The baseline donor pool contains the nineteen developing economies that, throughout 1996--2011, had diplomatic relations with the United States and with Israel and an institutional framework with the European Union in force, that faced no comprehensive UN, US or EU sanctions, experienced no coup, revolution or new civil war in 2010--13, and neither border Iran nor served as sanctions-evasion hubs. Iran's confrontation with the West is the treatment, so these \emph{normalisers} are the relevant comparison: countries in Iran's income range that took the opposite path. Online Appendix~\ref{A-app:donors} documents every inclusion and exclusion decision.

The observed common shock is the log price of Brent crude, and the exposure index is each country's mean oil rents in percent of GDP over 1996--2011 (WDI). Iran's value is 22.87. The donors range from essentially zero (Jordan, Kenya, the Dominican Republic, Panama) to 16.56 (Kazakhstan). Iran is therefore outside the exposure hull of the normalisers, with $d_{\min}=6.31$. Oil rents relative to GDP are a ratio of two outcomes and themselves respond to oil prices; the index is pre-determined only in the sense that it is measured before treatment, and Section~\ref{sec:app-robust} shows that a window ending in 2005, before the first UN measures, gives the same results. The \emph{exposure-feasible} pool adds Azerbaijan (26.52) and Angola (38.42). Both meet the diplomatic criteria and were excluded from the baseline pool because Azerbaijan borders Iran and Angola was at war until 2002. Adding them is a methodological choice, not a change in the population of comparators: in the exposure-feasible pool both receive zero synthetic-control weight, and their role is to place Iran inside the support of the exposure regression (Remark~\ref{rem:support}). Section~\ref{sec:app-robust} reports the results with each of them dropped.

\subsection{Diagnostics}\label{sec:app-diag}

\begin{figure}[htbp]\centering
\includegraphics[width=\textwidth]{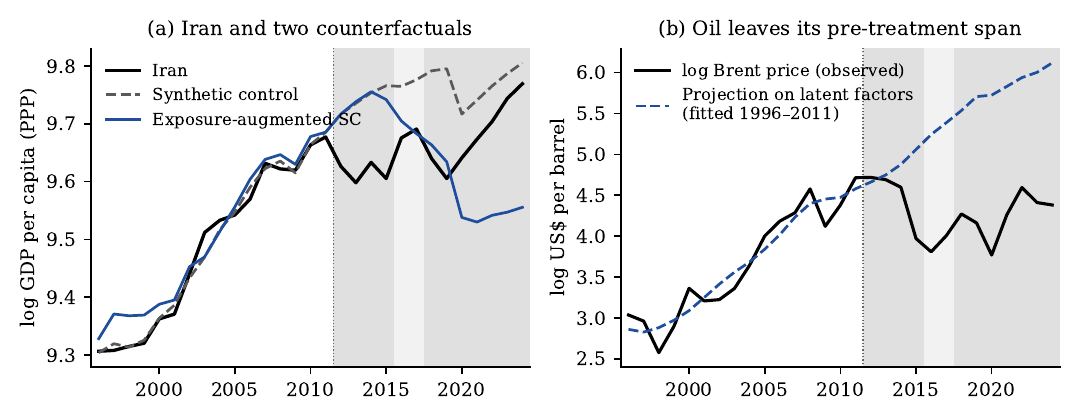}
\caption{Iran's GDP per capita and two counterfactuals (panel a) and the oil price against its projection on the donors' latent factors fitted over 1996--2011 (panel b). Shaded: 2012--15 shock and 2018--24 siege; light shading: 2016--17 relief.}
\label{fig:motivation}
\end{figure}

\begin{figure}[htbp]\centering
\includegraphics[width=\textwidth]{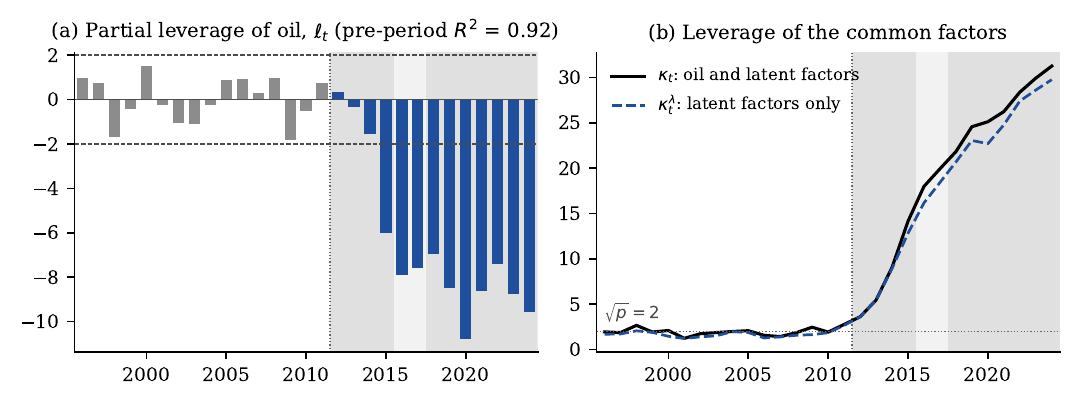}
\caption{Leverage diagnostics, normaliser pool, two exposure-orthogonalised latent factors. Panel (a): partial leverage $\hat\ell_t$ of log Brent on the latent factors (Corollary~\ref{cor:fwl}). Panel (b): leverage of the full factor vector, $\hat\kappa_t$, and of the latent factors alone, $\hat\kappa^\lambda_t$; the dotted line marks the pre-treatment root mean square $\sqrt p$.}
\label{fig:leverage}
\end{figure}

\begin{figure}[htbp]\centering
\includegraphics[width=0.78\textwidth]{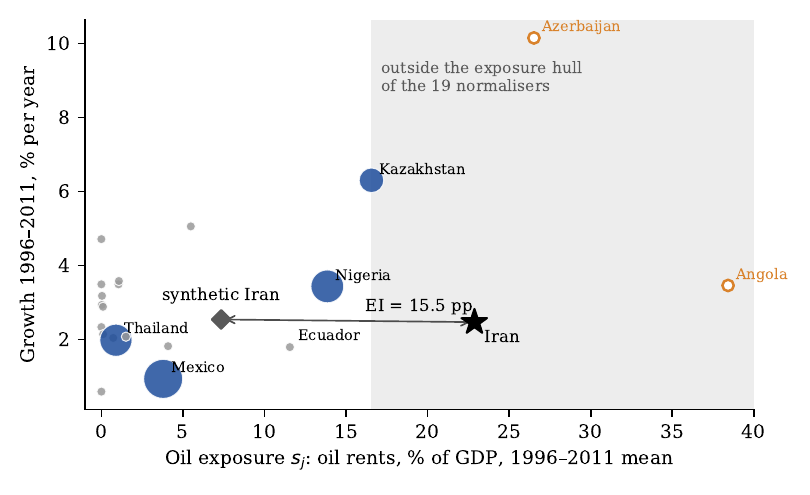}
\caption{Oil exposure and pre-treatment growth of the donors. Bubble area is proportional to the synthetic-control weight; grey: zero weight. Hollow markers: the two donors added in the exposure-feasible pool, which receive zero weight. The diamond is synthetic Iran.}
\label{fig:exposure}
\end{figure}

Table~\ref{tab:diag} and Figures~\ref{fig:motivation}--\ref{fig:exposure} report the diagnostics of Section~\ref{sec:diagnostics}. Before 2012, the two latent factors estimated from the donors track the oil price closely ($R^2=0.92$). Its partial leverage stays below 2 in absolute value in every pre-treatment year; from 2015 it falls to between $-6$ and $-11$ (Figure~\ref{fig:leverage}a). By the criterion of Corollary~\ref{cor:fwl}, oil is a silent factor: the boom of the 2000s moved with the donors' latent growth, and the collapse of 2014--16 did not.

The leverage of the full factor vector rises from about 3 in 2011 to 14 in 2015 and 31 in 2024 (Figure~\ref{fig:leverage}b). Most of this rise comes from the latent factors themselves, whose leverage reaches 30 by 2024, not from oil alone. The donors' latent growth paths after 2015 lie well outside their pre-treatment span. This warning applies to any synthetic control for Iran over this horizon, whatever is done about oil, and it is why the bias-aware intervals below widen sharply after 2015. The level of $\hat\kappa$ depends on how many latent factors are retained: with one, two and three factors the 2018--24 leverage in the normaliser pool is 12, 27 and 56 (Table~\ref{A-tab:factors-k}), and it is higher in the exposure-feasible pool, whose factors are estimated on a donor set that includes two economies with their own post-2014 collapses. The partial leverage of oil, which drives the exposure diagnosis, is less sensitive: its 2018--24 average ranges from $-9$ to $-15$ across these choices. I use two factors as the baseline, the smallest number that captures a trend and a cycle, and report breakdown values for one and three.

Figure~\ref{fig:exposure} shows the exposure geometry. Synthetic Iran is a combination of Mexico (weight 0.37), Nigeria (0.25), Thailand (0.24) and Kazakhstan (0.13). It reproduces Iran's average pre-treatment growth almost exactly, but its oil rents are 7.3 percent of GDP against Iran's 22.9: an exposure imbalance of 15.5 percentage points. Penalized SC, augmented SC, synthetic DiD and de-meaned SC have imbalances between 14.8 and 22.0 (Table~\ref{tab:estimators}). The imbalance is a feature of the pool, not of the estimator. Including oil rents among the predictors of a covariate synthetic control does not help either. The predictor weights chosen to minimise the pre-treatment MSPE give oil rents a weight that is numerically zero, and the imbalance is 13.3 points in the normaliser pool and 13.4 in the exposure-feasible pool. This is Corollary~\ref{cor:fwl} at work: pre-treatment outcomes give the data-driven predictor weights no reason to match on exposure.

\begin{table}[htbp]\centering\small
\caption{Diagnostics for out-of-span exposure}\label{tab:diag}
\begin{adjustbox}{max width=\textwidth}\begin{tabular}{@{}lcc@{}}\toprule
 & Normalisers (19) & Exposure-feasible (21)\\\midrule
\multicolumn{3}{@{}l}{\emph{Exposure geometry}}\\
\quad Treated exposure $s_1$ & 22.87 & 22.87\\
\quad Donor exposure range & [0.00, 16.56] & [0.00, 38.42]\\
\quad Distance to hull $d_{\min}$ & 6.31 & 0.00\\
\quad Exposure imbalance of SC, EI($\hat w$) & 15.53 & 15.53\\
\quad Pre-treatment RMSPE of SC & 0.0144 & 0.0144\\
\quad RMSPE at $d_{\min}$ (EBSC-min) & 0.3486 & 0.0526\\
\multicolumn{3}{@{}l}{\emph{Partial leverage of the oil price}}\\
\quad $R^2$ of oil on latent factors, 1996--2011 & 0.921 & 0.915\\
\quad $\max_{t\le 2011}|\ell_t|$ & 1.83 & 1.89\\
\quad $\max_{t\ge 2012}|\ell_t|$ & 10.76 & 8.22\\
\multicolumn{3}{@{}l}{\emph{Leverage of phase averages}}\\
\quad $\kappa$ (SC), 2012--2015 & 7.86 & 12.12\\
\quad $\kappa$ (SC), 2016--2017 & 18.95 & 28.12\\
\quad $\kappa$ (SC), 2018--2024 & 26.72 & 40.40\\
\quad $\kappa$ (EASC), 2012--2015 & 7.67 & 12.11\\
\quad $\kappa$ (EASC), 2016--2017 & 17.31 & 27.54\\
\quad $\kappa$ (EASC), 2018--2024 & 25.29 & 39.95\\
\bottomrule\end{tabular}\end{adjustbox}
\par\smallskip\begin{minipage}{0.92\textwidth}\footnotesize\emph{Notes.} Exposure is the 1996--2011 mean of oil rents in percent of GDP (WDI). EI is the treated unit's exposure minus the weighted donor average. Latent factors: two principal components of two-way demeaned donor outcomes after removing, period by period, the cross-sectional projection on exposure. $\ell_t$ is the residual of log Brent on a constant and the latent factors (fitted 1996--2011), divided by its pre-treatment root mean square (Corollary~\ref{cor:fwl}). $\kappa$ for a phase is the leverage of the phase-averaged factor vector (Proposition~\ref{prop:leverage}); the pre-treatment average of $\kappa_t^2$ equals the number of factors including the constant (4 for SC, 3 for the latent block). The EASC leverage is $(\kappa^{\lambda 2}+\ell_{2011}^2)^{1/2}$ (Proposition~\ref{prop:easc}).\end{minipage}
\end{table}

\begin{figure}[htbp]\centering
\includegraphics[width=0.62\textwidth]{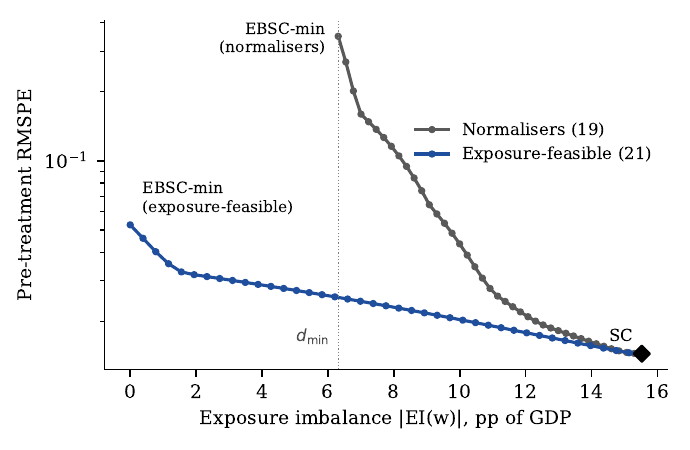}
\caption{Exposure--fit frontier $R(d)$: the smallest pre-treatment RMSPE attainable with simplex weights whose exposure imbalance is at most $d$. The left end of each curve is EBSC-min; the right end is synthetic control.}
\label{fig:frontier}
\end{figure}

The frontier in Figure~\ref{fig:frontier} shows what balance would cost. Within the normaliser pool, the smallest attainable imbalance, 6.3 points, requires putting all weight on Kazakhstan and raises the pre-treatment RMSPE from 0.014 to 0.35; no simplex weights both fit Iran and materially reduce its exposure gap. In the exposure-feasible pool exact balance is attainable at an RMSPE of 0.053. Along the frontier the RMSPE rises slowly until the imbalance falls to about 2 points and steeply thereafter, as the convexity result in Proposition~\ref{prop:geometry}(b) implies.

\subsection{Estimates}

Table~\ref{tab:main} reports the phase averages in log points. Standard synthetic control gives $-0.127$ in 2012--15, $-0.087$ in 2016--17 and $-0.089$ in 2018--24: a loss of about 12 percent that persists at about 8 percent through the relief and the siege. The exposure-augmented estimator agrees about the shock but not about what follows. With the exposure regression estimated on the exposure-feasible pool, so that Iran lies inside its support, EASC gives $-0.122$, $-0.011$ and $+0.109$. The tuning rule \eqref{eq:tuning} selects the synthetic-control weights in both pools, because the regression correction is cheaper than distorting the weights, so EASC$^*$ coincides with EASC. The exposure-balanced estimator in the exposure-feasible pool, which uses no regression correction, gives $-0.109$, $-0.022$ and $+0.054$. The two routes to exposure adjustment, reweighting and regression, agree that the persistent loss in the synthetic-control estimates is not robust. When the exposure regression uses only the 19 normalisers, the correction extrapolates beyond their exposure range and EASC gives $-0.152$, $-0.033$ and $+0.086$. The later-phase conclusion is unchanged; the shock estimate is larger and less reliable (Remark~\ref{rem:support} and Section~\ref{sec:app-robust}).

\begin{table}[htbp]\centering\small
\caption{Output effect of the 2012 sanctions: point estimates and inference}\label{tab:main}
\begin{adjustbox}{max width=\textwidth}\begin{tabular}{@{}lccccccc@{}}\toprule
 & & \multicolumn{3}{c}{Normalisers (19 donors)} & \multicolumn{3}{c}{Exposure-feasible (21 donors)}\\
\cmidrule(lr){3-5}\cmidrule(lr){6-8}
 & EI & 2012--15 & 2016--17 & 2018--24 & 2012--15 & 2016--17 & 2018--24\\\midrule
\multicolumn{8}{@{}l}{\emph{A. Point estimates (log points)}}\\
\quad SC & 15.5 & $-$0.127 & $-$0.087 & $-$0.089 & $-$0.127 & $-$0.087 & $-$0.089\\
\quad EBSC-min & 6.3 / 0.0 & \multicolumn{3}{c}{\emph{no adequate fit (RMSPE 0.349)}} & $-$0.109 & $-$0.022 & 0.054\\
\quad EASC & 15.5 & $-$0.152 & $-$0.033 & 0.086 & $-$0.122 & $-$0.011 & 0.109\\
\quad EASC$^*$ & 15.5 / 15.5 & $-$0.152 & $-$0.033 & 0.086 & $-$0.122 & $-$0.011 & 0.109\\
\quad Pre-treatment RMSPE: SC; EBSC-min &  & \multicolumn{3}{c}{0.0144; 0.3486} & \multicolumn{3}{c}{0.0144; 0.0526}\\
\multicolumn{8}{@{}l}{\emph{B. Inference for SC}}\\
\quad Placebo $p$-value &  & 0.050 & 0.300 & 0.350 & 0.045 & 0.364 & 0.500\\
\quad Noise s.d. $\hat\sigma$ &  & 0.049 & 0.111 & 0.180 & 0.048 & 0.111 & 0.181\\
\quad Leverage $\kappa$ &  & 7.9 & 18.9 & 26.7 & 12.1 & 28.1 & 40.4\\
\quad 95\% CI, $M=0$ &  & [$-$0.22, $-$0.03] & [$-$0.31, 0.13] & [$-$0.44, 0.26] & [$-$0.22, $-$0.03] & [$-$0.30, 0.13] & [$-$0.44, 0.27]\\
\quad 95\% CI, $M=0.5$ &  & [$-$0.26, 0.01] & [$-$0.41, 0.23] & [$-$0.58, 0.40] & [$-$0.29, 0.04] & [$-$0.47, 0.30] & [$-$0.68, 0.50]\\
\quad Breakdown $M^*$ &  & 0.41 & 0.00 & 0.00 & 0.27 & 0.00 & 0.00\\
\quad Breakdown $M^*$, robust $\hat\sigma$ &  & 0.55 & 0.00 & 0.00 & 0.35 & 0.00 & 0.00\\
\multicolumn{8}{@{}l}{\emph{C. Inference for EASC}}\\
\quad Placebo $p$-value &  & 0.050 & 0.550 & 0.300 & 0.091 & 0.909 & 0.273\\
\quad Noise s.d. $\hat\sigma$ &  & 0.064 & 0.150 & 0.244 & 0.052 & 0.117 & 0.183\\
\quad Leverage $\kappa$ &  & 7.7 & 17.3 & 25.3 & 12.1 & 27.5 & 39.9\\
\quad 95\% CI, $M=0$ &  & [$-$0.28, $-$0.03] & [$-$0.33, 0.26] & [$-$0.39, 0.56] & [$-$0.22, $-$0.02] & [$-$0.24, 0.22] & [$-$0.25, 0.47]\\
\quad 95\% CI, $M=0.5$ &  & [$-$0.31, 0.01] & [$-$0.41, 0.34] & [$-$0.50, 0.67] & [$-$0.29, 0.05] & [$-$0.40, 0.38] & [$-$0.48, 0.70]\\
\quad Breakdown $M^*$ &  & 0.41 & 0.00 & 0.00 & 0.21 & 0.00 & 0.00\\
\quad Breakdown $M^*$, robust $\hat\sigma$ &  & 0.49 & 0.00 & 0.00 & 0.27 & 0.00 & 0.00\\
\quad $100\times\bar\phi$ (jackknife $t$) &  & 0.16 (0.6) & $-$0.35 ($-$0.6) & $-$1.13 ($-$1.2) & $-$0.03 ($-$0.4) & $-$0.49 ($-$1.9) & $-$1.28 ($-$3.6)\\
\bottomrule\end{tabular}\end{adjustbox}
\par\smallskip\begin{minipage}{0.97\textwidth}\footnotesize\emph{Notes.} Entries are averages of annual gaps in log GDP per capita over each phase. EBSC-min: simplex weights with the smallest attainable exposure imbalance; EASC: SC weights plus the regression correction $\hat\phi_t\,\text{EI}(\hat w)$; EASC$^*$: frontier point minimising the estimated post-treatment MSE, then augmented. Placebo $p$-value: share of the $J+1$ units whose $|\bar\tau|/\text{RMSPE}$ is at least Iran's. $\hat\sigma$: dispersion of placebo phase averages standardised by each placebo's pre-treatment RMSPE and rescaled to Iran's; for EASC it adds $|\text{EI}|\,\overline{\text{se}}(\hat\phi)$ in quadrature. The exposure regression controls for each donor's growth within each third of 1996--2010. CIs are bias-aware (Proposition~\ref{prop:ci}) with bias bound $M\kappa\,\text{RMSPE}$; $M^*$ is the largest $M$ for which the CI excludes zero. Robust $\hat\sigma$ replaces the standard deviation by $1.4826\times$MAD. $\bar\phi$: phase average of the cross-donor exposure effect, log points per percentage point of oil rents, with delete-one-donor jackknife $t$-statistic.\end{minipage}
\end{table}

\begin{figure}[htbp]\centering
\includegraphics[width=0.66\textwidth]{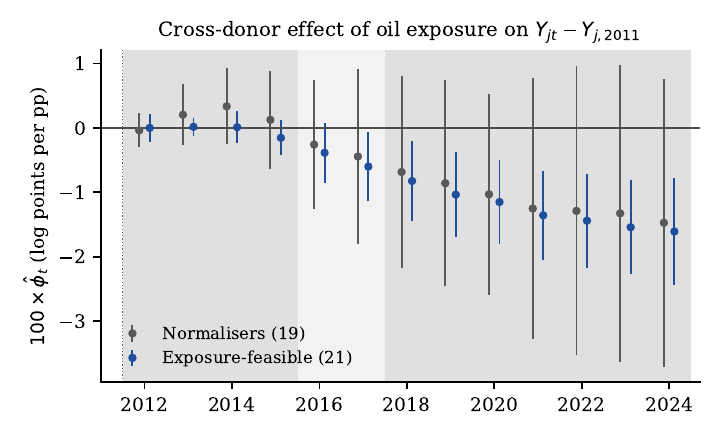}
\caption{Cross-donor exposure effects $\hat\phi_t$ with 95\% jackknife intervals: coefficient on oil rents in the regression of $Y_{jt}-Y_{j,2011}$ on oil rents and the donor's growth within each third of 1996--2010.}
\label{fig:phi}
\end{figure}

The cross-donor exposure effect $\hat\phi_t$ drives the correction (Figure~\ref{fig:phi}). It is close to zero through 2015, before the oil collapse had run its course, and increasingly negative from 2016. Over 2018--24, each percentage point of oil rents lowers the change in log GDP per capita since 2011 by 0.013 on average (jackknife $t=-3.6$ in the exposure-feasible pool). Multiplied by an exposure imbalance of 15.5 points, this accounts for 0.198 log points of the synthetic-control gap in 2018--24. Iran's output would have been expected to fall relative to its synthetic control even without sanctions, simply because Iran was far more exposed to oil. The regression has 21 observations and five parameters, and Angola and Azerbaijan have hat values of 0.70 and 0.61 (Figure~\ref{A-fig:influence}). The correction is therefore imprecise, and the inference below carries its jackknife variance. It is not, however, the product of one donor: deleting any single donor from the exposure-feasible pool leaves the 2018--24 coefficient between $-0.012$ and $-0.015$, and the normaliser pool alone, without either high-exposure donor, gives $-0.011$ with a wider interval. In that pool the coefficient is more fragile, ranging from $-0.005$ (without Nigeria) to $-0.016$ across deletions, one more reason to estimate the regression on a pool that brackets the treated unit.

\begin{figure}[htbp]\centering
\includegraphics[width=\textwidth]{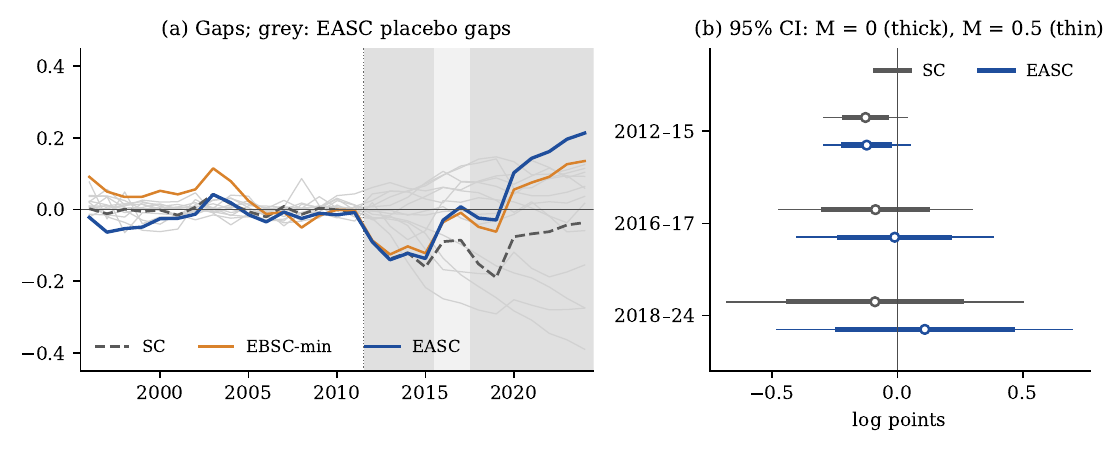}
\caption{Panel (a): annual gaps of synthetic control, EBSC-min and EASC (exposure-feasible pool), with EASC placebo gaps for donors whose pre-treatment MSPE is at most five times Iran's. Panel (b): phase averages with 95\% bias-aware intervals at $M=0$ (thick) and $M=0.5$ (thin).}
\label{fig:gaps}
\end{figure}

Figure~\ref{fig:gaps}(a) shows the annual paths. All three estimators open a gap of 0.12--0.14 log points by 2013. The synthetic-control gap stays between 0.085 and 0.19 through 2019. The exposure-adjusted estimators close the gap during the relief, reopen a small part of it in 2018--19, when US sanctions were reimposed and the waivers for buyers of Iranian oil ended, and turn positive from 2020. The positive gap after 2020 deserves the same scepticism as the negative gap it replaces. It lies inside the range of the placebo gaps in 2020 and somewhat above it from 2021, and the bias-aware interval for the phase includes zero. Three mechanisms could produce a spurious positive estimate. A linear correction over-corrects when exposure is concave in oil rents, which design S5 of the simulations illustrates and the resource-curse literature makes plausible \citep{vanderploeg2011}. A recovery in Iran's oil exports, mostly to China \citep{eia2026}, raised its effective exposure above the pre-treatment index. And official growth may be overstated \citep{martinez2022}. The V-Dem exposure index gives a 2018--24 estimate of about zero (Table~\ref{tab:robust}). The defensible reading is that the long-run effect is not identified.

\subsection{Standard estimators before and after exposure adjustment}\label{sec:app-estimators}

\begin{figure}[htbp]\centering
\includegraphics[width=\textwidth]{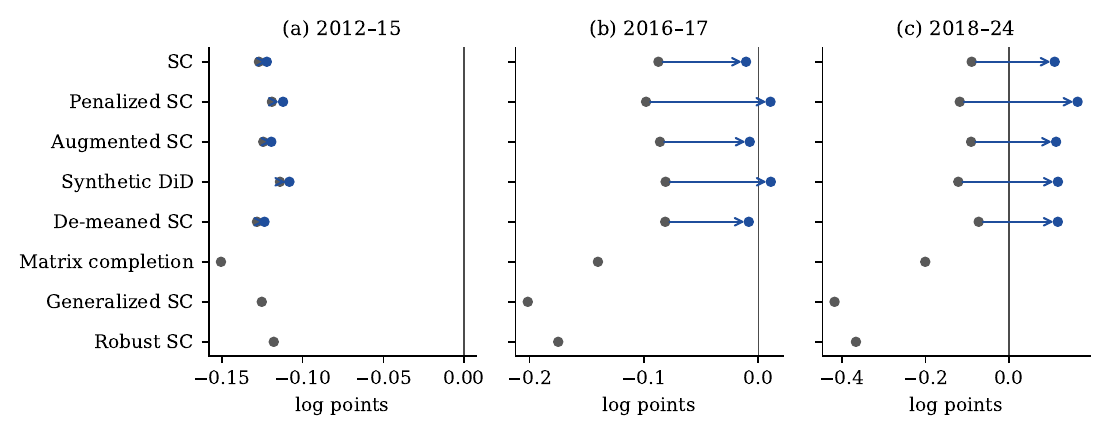}
\caption{Phase averages of standard estimators as published (grey) and after the exposure correction $\hat\phi_t\,\EI(\hat w)$ (blue), exposure-feasible pool. Estimators without weights are shown as published.}
\label{fig:estimators}
\end{figure}

Table~\ref{tab:estimators} and Figure~\ref{fig:estimators} apply the same correction to the weights of every standard estimator; Table~\ref{A-tab:estimators-base} in the Online Appendix repeats the exercise for the normaliser pool. As published, the estimators disagree about the later phases. Penalized SC and synthetic DiD find losses of 0.08--0.12 log points, de-meaned SC and synthetic control 0.07--0.09, and synthetic control with covariates 0.05--0.06. Once each estimator's own exposure imbalance is corrected with the common $\hat\phi_t$, they agree closely: $-0.11$ to $-0.12$ in 2012--15, $-0.01$ to $+0.01$ in 2016--17 and $+0.11$ to $+0.16$ in 2018--24. Most of the disagreement among the standard estimators is disagreement about how much oil exposure to give synthetic Iran.

\begin{table}[htbp]\centering\small
\caption{Exposure imbalance of standard estimators and their exposure-augmented versions}\label{tab:estimators}
\begin{adjustbox}{max width=\textwidth}\begin{tabular}{@{}lccccccc@{}}\toprule
 & & \multicolumn{3}{c}{As published} & \multicolumn{3}{c}{Exposure-augmented}\\
\cmidrule(lr){3-5}\cmidrule(lr){6-8}
Estimator & EI($\hat w$) & 2012--15 & 2016--17 & 2018--24 & 2012--15 & 2016--17 & 2018--24\\\midrule
SC & 15.5 & $-$0.127 & $-$0.087 & $-$0.089 & $-$0.122 & $-$0.011 & 0.109\\
SC with covariates, oil rents a predictor & 13.4 & $-$0.116 & $-$0.061 & $-$0.047 & $-$0.112 & 0.005 & 0.125\\
Penalized SC & 22.0 & $-$0.119 & $-$0.098 & $-$0.118 & $-$0.112 & 0.011 & 0.164\\
Augmented SC & 15.9 & $-$0.124 & $-$0.086 & $-$0.091 & $-$0.119 & $-$0.008 & 0.113\\
Synthetic DiD & 18.6 & $-$0.114 & $-$0.081 & $-$0.121 & $-$0.108 & 0.011 & 0.117\\
De-meaned SC & 14.8 & $-$0.128 & $-$0.081 & $-$0.073 & $-$0.124 & $-$0.008 & 0.117\\
Matrix completion & -- & $-$0.151 & $-$0.140 & $-$0.201 & -- & -- & --\\
Generalized SC & -- & $-$0.125 & $-$0.201 & $-$0.418 & -- & -- & --\\
Robust SC & -- & $-$0.118 & $-$0.175 & $-$0.366 & -- & -- & --\\
\bottomrule\end{tabular}\end{adjustbox}
\par\smallskip\begin{minipage}{0.95\textwidth}\footnotesize\emph{Notes.} Exposure-feasible pool (21 donors); the correction $\hat\phi_t$ is the same for every estimator. EI is defined for estimators that produce donor weights summing to one (augmented SC weights may be negative). SC with covariates: nested-$V$ synthetic control with predictors oil rents, outcome means over 1996--2000, 2001--05 and 2006--11, and the 2011 outcome; the selected $V$ gives oil rents a weight below $10^{-11}$. Matrix completion, generalized and robust SC produce no weights; their counterfactuals extrapolate Iran's loadings (Section~\ref{sec:app-estimators}). Tuning parameters are chosen by forecasting Iran over 2008--11 from 1996--2007.\end{minipage}
\end{table}


The estimators that produce no weights behave as Proposition~\ref{prop:leverage} predicts for regression-based extrapolation under high leverage. In the normaliser pool, generalized SC and robust SC give $-0.03$ and $-0.04$ for the 2012--15 shock and gains of 0.22--0.27 log points in 2018--24. Adding Azerbaijan and Angola changes these to $-0.13$ and $-0.12$ and losses of 0.37--0.42 log points. Their counterfactuals extrapolate Iran's loading on a factor that is nearly collinear with the trend before 2012; that loading is weakly identified, and its prediction variance scales with $\kappa_t^2$. Matrix completion, which shrinks toward a low-rank fit, is more stable but inherits the oil imbalance of the donors it implicitly weights ($-0.20$ in 2018--24).

\subsection{Inference and sensitivity}\label{sec:app-inference}

The 2012--15 shock is the only phase effect distinguishable from zero, and even there the evidence is exactly as strong as 20 placebo units allow. For synthetic control, Iran's ratio $|\bar\tau|/\RMSPE$ is the largest among the 20 (22) units in the normaliser (exposure-feasible) pool, so the placebo $p$-values are the minimum attainable, 0.05 and 0.045. The bias-aware interval at $M=0$ is $[-0.22,-0.03]$ in both pools. For EASC in the exposure-feasible pool, Iran ranks second (placebo $p=0.09$) and the interval at $M=0$ is $[-0.22,-0.02]$. For the later phases, all intervals include zero and the placebo $p$-values range from 0.27 to 0.91.

\begin{figure}[htbp]\centering
\includegraphics[width=\textwidth]{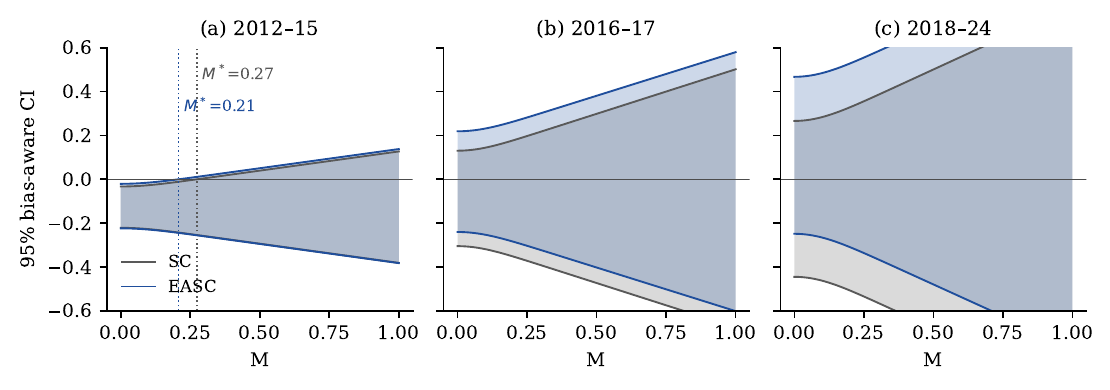}
\caption{95\% bias-aware confidence intervals as a function of the sensitivity parameter $M$ (share of pre-treatment RMSPE treated as systematic), exposure-feasible pool. $M^*$: breakdown value.}
\label{fig:sensitivity}
\end{figure}

How robust is the shock estimate to systematic misfit? The breakdown values in Table~\ref{tab:main} and Figure~\ref{fig:sensitivity} are 0.27--0.41 for synthetic control and 0.21--0.41 for EASC, or 0.27--0.55 with the robust noise scale. The conclusion that the 2012--15 effect is negative survives if at most a fifth to two-fifths of the pre-treatment RMSPE of 0.014 is systematic and aligned with the post-treatment factor direction; a larger share would overturn it. Given the leverage of 8--12 in this phase, that is moderate robustness, not overwhelming robustness. The breakdown values depend on the number of latent factors through $\kappa$: with one factor they rise to 0.96 (normalisers) and 1.5 (exposure-feasible), with three they fall to 0.30 and 0.23 (Table~\ref{A-tab:factors-k}). For the later phases the leverage is 17--40 with two factors and 6--12 with one, so even modest systematic misfit swamps the estimate. With these donors and this horizon, the data cannot settle whether the sanctions left a lasting output loss. Leverage diagnostics make this explicit; conventional placebo inference does not.

\subsection{Robustness}\label{sec:app-robust}

Table~\ref{tab:robust} varies the exposure measure, the controls in the exposure regression and the pool. Across ten specifications, EASC ranges from $-0.153$ to $-0.122$ in 2012--15, from $-0.060$ to $0.016$ in 2016--17 and from $-0.015$ to $0.138$ in 2018--24. The exposure-balanced estimator, where feasible, gives similar answers. No specification reproduces the persistent loss of about 0.09 log points that synthetic control reports. Four deserve comment. With the V-Dem measure of petroleum, coal and gas income \citep{haber2011} as the exposure index, the 2018--24 effect is close to zero rather than positive. Without pre-treatment growth controls in the exposure regression, the correction is smaller. Proposition~\ref{prop:easc}(a) shows why the controls matter: without them, $\hat\phi_t$ also absorbs any correlation between exposure and composite latent loadings. With Angola as the only addition to the normalisers, so that no donor borders Iran, the results are close to those of the full exposure-feasible pool; with Azerbaijan as the only addition, so that Angola, the most influential observation in the exposure regression, is absent, the 2018--24 estimate rises to $0.138$ and the shock estimate is unchanged. Measuring exposure over 1996--2005, before the first UN measures and the 2006--08 price spike, changes nothing. EASC at the synthetic-control weights does not depend on the number of latent factors, which enters only the tuning rule and the inference. With one factor, EASC$^*$ in the normaliser pool moves to a frontier point with imbalance 12.5 and gives $-0.162$, $-0.048$ and $+0.053$; the inference consequences of the number of factors are in Table~\ref{A-tab:factors-k}.

\begin{table}[htbp]\centering\small
\caption{Robustness of the exposure-augmented estimates}\label{tab:robust}
\begin{adjustbox}{max width=\textwidth}\begin{tabular}{@{}lccccccc@{}}\toprule
 & & \multicolumn{3}{c}{EASC} & \multicolumn{3}{c}{EBSC-min}\\
\cmidrule(lr){3-5}\cmidrule(lr){6-8}
Specification & EI(SC) & 2012--15 & 2016--17 & 2018--24 & 2012--15 & 2016--17 & 2018--24\\\midrule
Oil rents 1996--2011 (baseline) & 15.5 & $-$0.152 & $-$0.033 & 0.086 & \emph{infeasible} &  & \\
Oil rents 2002--2011 & 16.8 & $-$0.153 & $-$0.040 & 0.069 & \emph{infeasible} &  & \\
V-Dem oil income share & 1.6 & $-$0.137 & $-$0.058 & 0.002 & \emph{infeasible} &  & \\
No pre-treatment growth controls in the exposure regression & 15.5 & $-$0.146 & $-$0.057 & 0.004 & \emph{infeasible} &  & \\
Exposure-feasible pool (21) & 15.5 & $-$0.122 & $-$0.011 & 0.109 & $-$0.109 & $-$0.022 & 0.054\\
Exposure-feasible pool, V-Dem share & 1.6 & $-$0.125 & $-$0.060 & $-$0.015 & $-$0.115 & $-$0.056 & $-$0.026\\
Exposure-feasible pool without Azerbaijan (20) & 15.5 & $-$0.126 & $-$0.018 & 0.101 & $-$0.109 & $-$0.022 & 0.054\\
Exposure-feasible pool without Angola (20) & 15.5 & $-$0.123 & 0.016 & 0.138 & \emph{infeasible} &  & \\
Exposure-feasible pool, oil rents 1996--2005 & 14.9 & $-$0.123 & $-$0.012 & 0.109 & $-$0.107 & $-$0.020 & 0.055\\
Exposure-feasible pool, no growth controls & 15.5 & $-$0.125 & $-$0.031 & 0.041 & $-$0.109 & $-$0.022 & 0.054\\
\midrule
SC (all specifications) &  & $-$0.127 & $-$0.087 & $-$0.089 &  &  & \\
\bottomrule\end{tabular}\end{adjustbox}
\par\smallskip\begin{minipage}{0.97\textwidth}\footnotesize\emph{Notes.} Rows 1--4 use the 19 normalisers; rows 5--10 add Azerbaijan and Angola (row 7 adds Angola only, row 8 Azerbaijan only). The V-Dem measure is petroleum, coal and gas income per capita relative to GDP per capita \citep{haber2011}, 1996--2006 mean; its scale differs from oil rents, so EI is not comparable across measures. EBSC-min is reported as infeasible when the minimum attainable imbalance forces a pre-treatment RMSPE above 0.1. SC estimates do not depend on the exposure measure or on the latent factors.\end{minipage}
\end{table}


The Online Appendix reports three further exercises. In-time placebos with fictitious treatment dates between 2006 and 2009, on data ending in 2011 (Table~\ref{A-tab:intime}), bracket the 2008--09 oil collapse. Relative to the pre-treatment span, that collapse was smaller and shorter than the one of 2014--16 ($\max|\hat\ell_t|\le2.6$ in the normaliser pool). Mean placebo gaps are at most 0.07 for synthetic control and 0.05 for EASC in either pool, with one exception: for the 2006 date in the exposure-feasible pool, EASC has a gap of $-0.10$ against $-0.07$ for synthetic control. That placebo window contains the UN sanctions of 2006--10, and its pre-period is only ten years long. Backdating the first treated year to 2009, 2010 or 2011 (Table~\ref{A-tab:backdate}) leaves synthetic control ($-0.13$) and EASC in the exposure-feasible pool ($-0.10$ to $-0.11$) essentially unchanged for 2012--15; in the normaliser pool EASC moves to between $-0.18$ and $-0.27$, because there its correction extrapolates. Dropping each donor with positive weight in turn (Table~\ref{A-tab:loo}) moves the synthetic-control shock between $-0.107$ and $-0.145$, and the EASC shock between $-0.101$ and $-0.139$ in the exposure-feasible pool and between $-0.138$ and $-0.194$ in the normaliser pool. The EASC estimate for 2018--24 ranges from 0.04 to 0.22 in the exposure-feasible pool and from $-0.10$ to $0.20$ in the normaliser pool, confirming that the later phases are imprecisely estimated.

\subsection{Interpretation}

Three conclusions follow. First, the 2012 package of financial disconnection and oil embargo caused a sharp output loss of about 0.12--0.13 log points, or 11--12 percent of GDP per capita, within four years. Every weighting estimator that fits the pre-treatment period finds a loss of this order, between 0.11 and 0.15 log points, in both donor pools and with or without the exposure adjustment; backdating changes it little except where the exposure correction extrapolates. The bias-aware interval excludes zero unless more than a fifth to two-fifths of the pre-treatment misfit is systematic. The magnitude is in line with the oil-export channel emphasised by \citet{laudati2023}: Iranian crude and condensate exports fell from 2.6 million barrels a day in 2011 to just under 1.3 million in 2013 \citep{eia2015}.

Second, the persistent loss that standard synthetic control attributes to sanctions after 2016 is not robust. It coincides with a collapse in the price of the commodity on which Iran depended far more than any comparator, and any adjustment for that exposure, by reweighting or by regression and with either exposure index, removes it. The adjusted estimates for 2018--24 range from about zero to $+0.14$ and are themselves imprecise; they establish no gain. Earlier synthetic-control studies of the Iranian sanctions found large and persistent effects \citep{gharehgozli2017}, but none of them reported the exposure imbalance of their synthetic Iran. The diagnostics in Section~\ref{sec:diagnostics} would have flagged it.

Third, absence of evidence of a persistent loss is not evidence of absence. Latent leverage after 2015 is so high that no estimator in this class can pin down the long-run effect with these comparators. Mechanisms that could produce a lasting loss, such as the expansion of Revolutionary Guard business networks into smuggling and import substitution, the reallocation of rents to politically connected firms, and weakened property rights \citep{acemoglu2001,mehlum2006,vanderploeg2011}, could be offset in the aggregate by exchange-rate depreciation and import substitution. The aggregate data cannot distinguish these possibilities.

\section{Conclusion and practical recommendations}\label{sec:conclusion}

Synthetic control trusts pre-treatment fit to balance whatever drives untreated outcomes. This paper has identified a class of cases in which that trust is misplaced, and shown that the class can be diagnosed with information available before the post-treatment outcomes are examined. An observed common shock that moves with the latent factors before treatment and breaks away from them afterwards is a silent factor. Pre-treatment fit does not restrict the synthetic unit's exposure to it, and the resulting bias grows with the shock's post-treatment partial leverage. The problem is not peculiar to synthetic control: every simplex-weighted estimator inherits it when the treated unit lies outside the donors' exposure hull, and regression-based estimators pay for the same extrapolation in variance. The remedy combines balance where it is feasible with a regression correction where it is not. The correction is a transparent adjustment resting on a stated assumption about the donors' cross-section, not an identified estimator in its own right. It should be estimated on donors whose exposure brackets the treated unit's, whether or not they receive weight, reported with its influence diagnostics, and read alongside the balanced estimator where one exists.

For applied work I recommend five steps, all implemented in the replication package.
\begin{enumerate}[itemsep=1pt]
\item Identify the observed common shocks to which the treated unit is unusually exposed (commodity prices, exchange rates, partner demand), and plot their partial leverage $\hat\ell_t$ with respect to the donors' latent factors. A pre-treatment $R^2$ near one and post-treatment $|\hat\ell_t|$ well above 2 indicate a silent factor.
\item Report the exposure imbalance $\EI(\hat w)$ of every estimator, the distance to the hull $d_{\min}$, and the exposure--fit frontier.
\item If the exposure effect $\hat\phi_t$ is non-negligible, report the exposure-augmented estimator alongside the standard one, with the hat values and delete-one estimates of the exposure regression. If the treated unit is outside the hull, enlarge the pool used for the exposure regression until its support covers the treated unit, and report the balanced estimator as well.
\item Report the leverage $\hat\kappa$ of each post-treatment phase and bias-aware intervals indexed by $M$, together with the breakdown value $M^*$. At long horizons, latent leverage alone can make conclusions fragile, whatever is done about observed shocks.
\item Treat factor-model estimators that extrapolate loadings with caution when leverage is high; instability across reasonable donor pools is the symptom.
\end{enumerate}

The analysis has limitations that point to extensions. The correction can remove only the part of the post-treatment divergence that exposure predicts across donors; a treated unit that loads idiosyncratically on an out-of-span factor, as in design S11, lies beyond the reach of any method in this class, and the diagnostics cannot detect it. Exposure is assumed linear in a pre-determined index. Nonlinear exposure can be handled with a local or series regression in place of \eqref{eq:phi-reg}, at the cost of a stronger support requirement. The index must not respond to treatment; oil rents measured before 2012 satisfy this by construction, but contemporaneous measures would not. The sensitivity parameter $M$ is not identified from the data; it is a device for reporting how much systematic misfit a conclusion can tolerate. Values up to 1 are natural, but larger values remain possible when the weights over-fit the pre-treatment period. Finally, the framework extends directly to several treated units and staggered adoption \citep{benmichael2022}, where exposure imbalance can be computed cohort by cohort.

For Iran, the conclusion is specific. Financial disconnection and the oil embargo cut output sharply, by about 12 percent within four years. Whether they also lowered output permanently cannot be established with synthetic control and the available comparators. The persistent loss that standard estimates suggest reflects Iran's oil dependence, not an identified effect of sanctions.

\paragraph{Data availability.} All data are publicly available: World Development Indicators (World Bank), the Brent price (FRED series POILBREUSDA) and V-Dem v16. The replication package, deposited in the \emph{Journal of Applied Econometrics} Data Archive, contains the extracts used, scripts that download fresh vintages, and code that reproduces every table, figure and number in the paper and the Online Appendix.

\clearpage
\bibliographystyle{plainnat}
\bibliography{refs}

\end{document}


\maketitle
\tableofcontents
\clearpage

\section{Proofs}\label{app:proofs}

Throughout, $w$ is any weight vector with $\sum_{j\ge2}w_j=1$, $\beta_j=(\mu_j',\theta_j')'$, $g_t=(\lambda_t',x_t')'$, and all statements except those of Proposition~\ref{M-prop:easc}(a) and (c), which treat loadings as random given $(s_j,c_j)$, are conditional on factors and loadings.

\paragraph{Proof of Lemma~\ref{M-lem:decomp}.} For $t\le T_0$ all outcomes are untreated, so by \eqref{M-eq:model}
\[
e_t(w)=\delta_t\Big(1-\sum_jw_j\Big)+\Big(\mu_1-\sum_jw_j\mu_j\Big)'\lambda_t+\Big(\theta_1-\sum_jw_j\theta_j\Big)'x_t+\varepsilon_{1t}-\sum_jw_j\varepsilon_{jt}=\Delta(w)'g_t+u_t(w).
\]
For $t>T_0$, $Y_{1t}=Y^N_{1t}+\tau_t$ and donors are untreated, so the same algebra gives $\hat\tau_t(w)=\tau_t+\Delta(w)'g_t+u_t(w)$. \qed

\paragraph{Proof of Proposition~\ref{M-prop:leverage}.} By Lemma~\ref{M-lem:decomp}, $\hat\tau_t(w)-\tau_t=\Delta'g_t+u_t(w)$; if $w$ is fixed or independent of the post-treatment noise, $\E[u_t(w)\mid\beta,g]=0$ and $\Delta'g_t$ is the conditional bias. The inequality itself is algebraic. Since $S_0$ is symmetric positive definite,
\[
|\Delta'g_t|=|(S_0^{1/2}\Delta)'(S_0^{-1/2}g_t)|\le\|S_0^{1/2}\Delta\|\,\|S_0^{-1/2}g_t\|=\eta(w)\,\kappa_t
\]
by the Cauchy--Schwarz inequality. Equality holds for $\Delta=\eta\,S_0^{-1}g_t/\kappa_t$, for which $\Delta'S_0\Delta=\eta^2 g_t'S_0^{-1}g_t/\kappa_t^2=\eta^2$ and $\Delta'g_t=\eta\kappa_t$. For an average over $\mathcal T$, the bias is $\Delta'\bar g_{\mathcal T}$ and the same argument applies with $\bar g_{\mathcal T}$; $\kappa_{\mathcal T}=\|S_0^{-1/2}\bar g_{\mathcal T}\|\le|\mathcal T|^{-1}\sum_{t\in\mathcal T}\|S_0^{-1/2}g_t\|$ by the triangle inequality. Next, $T_0^{-1}\sum_{t\le T_0}g_t'S_0^{-1}g_t=\mathrm{tr}\big(S_0^{-1}T_0^{-1}\sum_{t\le T_0}g_tg_t'\big)=\mathrm{tr}(I_p)=p$. Finally, let $G_0$ be the $T_0\times p$ matrix with rows $g_t'$, $t\le T_0$. Then $G_0'G_0=T_0S_0$, the prediction-variance factor of a point $g_t$ in the least-squares regression on $G_0$ is $g_t'(G_0'G_0)^{-1}g_t=\kappa_t^2/T_0$, and the variance of the fitted value $g_t'\hat c$ under homoskedastic noise with variance $\sigma^2$ is $\sigma^2\kappa_t^2/T_0$. \qed

\paragraph{Proof of Corollary~\ref{M-cor:fwl}.} Substituting $x_t=A'\lambda_t+r_t$ into \eqref{M-eq:model} gives $\mu_j'\lambda_t+\theta_j'x_t=(\mu_j+A\theta_j)'\lambda_t+\theta_j'r_t=\tilde\mu_j'\lambda_t+\theta_j'r_t$, which is the stated representation. By construction of $A$, $\sum_{t\le T_0}\lambda_tr_t'=0$.

(a) Partition $S_0$ conformably with $(\lambda_t,x_t)$:
\[
S_0=\begin{pmatrix}S_\lambda&S_{\lambda x}\\S_{x\lambda}&S_x\end{pmatrix},\qquad A=S_\lambda^{-1}S_{\lambda x},\qquad S_x-S_{x\lambda}S_\lambda^{-1}S_{\lambda x}=T_0^{-1}\sum_{t\le T_0}r_tr_t'=S_r .
\]
The partitioned-inverse formula gives, for any $g=(\lambda',x')'$,
\[
g'S_0^{-1}g=\lambda'S_\lambda^{-1}\lambda+(x-A'\lambda)'S_r^{-1}(x-A'\lambda),
\]
so $\kappa_t^2=(\kappa^\lambda_t)^2+r_t'S_r^{-1}r_t=(\kappa^\lambda_t)^2+\|\ell_t\|^2$.

(b) $\Delta'g_t=\Delta_\mu'\lambda_t+\Delta_\theta'x_t=(\Delta_\mu+A\Delta_\theta)'\lambda_t+\Delta_\theta'r_t=\tilde\Delta'\lambda_t+\Delta_\theta'r_t$. Squaring and averaging over $t\le T_0$, the cross term $2\tilde\Delta'(T_0^{-1}\sum_{t\le T_0}\lambda_tr_t')\Delta_\theta$ vanishes, which gives $\eta^2=\tilde\Delta'S_\lambda\tilde\Delta+\Delta_\theta'S_r\Delta_\theta$.

(c) From (b), $\eta^2\ge\Delta_\theta^2S_r$ when $q=1$, so $\eta\le\bar\eta$ implies $|\Delta_\theta|\le\bar\eta/S_r^{1/2}$. If $\tilde\Delta=0$, then $\eta=|\Delta_\theta|S_r^{1/2}$ and the bias is $\Delta_\theta r_t$, so their ratio is $|r_t|/S_r^{1/2}=|\ell_t|$. \qed

\paragraph{Proof of Proposition~\ref{M-prop:balance}.} Under Assumption~\ref{M-ass:index}, $\Delta_\theta(w)=\Gamma s_1-\sum_jw_j\Gamma s_j=\Gamma\,\EI(w)$. If $\EI(w)=0$, then $\Delta_\theta=0$, and Corollary~\ref{M-cor:fwl}(b) gives $\Delta'g_t=\tilde\Delta'\lambda_t$ and $\eta^2=\tilde\Delta'S_\lambda\tilde\Delta$. The Cauchy--Schwarz argument of Proposition~\ref{M-prop:leverage}, applied in the metric of $S_\lambda$, gives $|\tilde\Delta'\lambda_t|\le\kappa^\lambda_t\eta$. \qed

\paragraph{Proof of Proposition~\ref{M-prop:geometry}.} (a) The map $w\mapsto\sum_jw_js_j$ sends $\Delta^J$ onto $\conv\mathcal S$. Hence $\{\EI(w):w\in\Delta^J\}=s_1-\conv\mathcal S$ and $\min_w\|\EI(w)\|=\dist(s_1,\conv\mathcal S)$, which is zero if and only if $s_1\in\conv\mathcal S$ because $\conv\mathcal S$ is closed. For scalar exposure $\conv\mathcal S=[\min_js_j,\max_js_j]$.

(b) Let $f(w)=\RMSPE(w)=\|T_0^{-1/2}(y_1-Y_0w)\|$, a convex function of $w$ (a norm of an affine map). The feasible set $\{w\in\Delta^J:\|\EI(w)\|\le d\}$ grows with $d$, so $R$ is non-increasing. For convexity, take $d_1,d_2\ge d_{\min}$, minimisers $w_1,w_2$ and $\alpha\in[0,1]$. The point $w_\alpha=\alpha w_1+(1-\alpha)w_2$ lies in $\Delta^J$, and since $\EI$ is affine, $\|\EI(w_\alpha)\|\le\alpha\|\EI(w_1)\|+(1-\alpha)\|\EI(w_2)\|\le\alpha d_1+(1-\alpha)d_2$. Hence $R(\alpha d_1+(1-\alpha)d_2)\le f(w_\alpha)\le\alpha f(w_1)+(1-\alpha)f(w_2)=\alpha R(d_1)+(1-\alpha)R(d_2)$.

(c) By Lemma~\ref{M-lem:decomp} and Corollary~\ref{M-cor:fwl}(b) with $\Delta_\theta=\gamma\EI(w)$, $\hat\tau_t(w)-\tau_t=\tilde\Delta(w)'\lambda_t+\gamma\EI(w)r_t+u_t(w)$, and $|\EI(w)|\ge d_{\min}$ by (a). For an estimator that subtracts a (weighted) pre-treatment mean, $\hat\tau_t=[Y_{1t}-\sum_s\pi_sY_{1s}]-\sum_jw_j[Y_{jt}-\sum_s\pi_sY_{js}]$ with $\sum_{s\le T_0}\pi_s=1$; the same algebra applies to $g_t-\sum_s\pi_sg_s$, so $r_t$ is replaced by $r_t-\sum_s\pi_sr_s$. \qed

\paragraph{Proof of Proposition~\ref{M-prop:easc}.} (a) For donors, by Corollary~\ref{M-cor:fwl} and Assumption~\ref{M-ass:index},
\[
Y_{jt}-Y_{jb}=(\delta_t-\delta_b)+\tilde\mu_j'(\lambda_t-\lambda_b)+s_j'\Gamma'(r_t-r_b)+(\varepsilon_{jt}-\varepsilon_{jb}).
\]
Since the first element of $\lambda_t$ is constant, $\tilde\mu_j'(\lambda_t-\lambda_b)=\tilde\mu^{-\prime}_j(\lambda^{-}_t-\lambda^{-}_b)$. Under Assumption~\ref{M-ass:span},
\[
\E[Y_{jt}-Y_{jb}\mid s_j,c_j]=\big[\delta_t-\delta_b+\alpha_0'(\lambda^{-}_t-\lambda^{-}_b)\big]+s_j'\Gamma'(r_t-r_b)+c_j'\Pi'(\lambda^{-}_t-\lambda^{-}_b),
\]
which is linear in $(1,s_j,c_j)$. The population least-squares projection therefore coincides with the conditional expectation, and its coefficient on $s_j$ is $\phi_t=\Gamma'(r_t-r_b)$.

(b) By Lemma~\ref{M-lem:decomp} and Corollary~\ref{M-cor:fwl}(b), $\hat\tau_t(w)-\tau_t=\tilde\Delta'\lambda_t+\EI(w)'\Gamma'r_t+u_t(w)$. Writing $\Gamma'r_t=\Gamma'(r_t-r_b)+\Gamma'r_b=\phi_t+\Gamma'r_b$ and subtracting $\hat\phi_t'\EI(w)$ gives \eqref{M-eq:easc-error}.

(c) The exposure component of the synthetic-control bias is $\EI(w)'\Gamma'(r_t-r_b)=\EI(w)'\phi_t$; after correction it is $\EI(w)'(\phi_t-\hat\phi_t)$. For fixed $w$, its conditional expectation given $\mathcal C=\{s_j,c_j\}_{j\ge2}$ is $\EI(w)'(\phi_t-\E[\hat\phi_t\mid\mathcal C])$, which is zero identically if $\EI(w)=0$. Under Assumption~\ref{M-ass:span}, whose cross-donor mean-independence condition gives $\E[Y_{jt}-Y_{jb}\mid\mathcal C]=\E[Y_{jt}-Y_{jb}\mid s_j,c_j]$, $\E[Y_{jt}-Y_{jb}\mid\mathcal C]$ is linear in $(1,s_j,c_j)$ with coefficient $\phi_t$ on $s_j$ by part (a), so least squares is conditionally unbiased, $\E[\hat\phi_t\mid\mathcal C]=\phi_t$, and the term is again zero. With estimated weights, $\hat w$ and $\hat\phi_t$ share donors' pre-treatment noise, so this argument holds only approximately; the simulations in Section~\ref{M-sec:mc} include this dependence.

(d) Let $a=(S_\lambda^{1/2}\tilde\Delta;\,S_r^{1/2}\Gamma\EI(w))$ and $v=(S_\lambda^{-1/2}\lambda_t;\,S_r^{-1/2}r_b)$. Then $a'v=\tilde\Delta'\lambda_t+\EI(w)'\Gamma'r_b$, $\|a\|^2=\tilde\Delta'S_\lambda\tilde\Delta+\EI'\Gamma'S_r\Gamma\EI=\eta(w)^2$ by Corollary~\ref{M-cor:fwl}(b), and $\|v\|^2=(\kappa^\lambda_t)^2+\|\ell_b\|^2$. The Cauchy--Schwarz inequality completes the proof. \qed

\begin{remark}[When the controls also load on exposure]\label{rem:approx}
Let the controls be sub-period changes, $c_j=\big(Y_{j,e_k}-Y_{j,s_k}\big)_{k=1}^K$ with $s_k<e_k<b$. By Corollary~\ref{M-cor:fwl}, $c_j=\Delta\delta^c+L'\tilde\mu^{-}_j+R_c'\Gamma s_j+\Delta\varepsilon^c_j$, where the columns of $L$ ($(r-1)\times K$) and $R_c$ ($q\times K$) are the sub-period changes of the non-constant latent factors $\lambda^{-}_t$ and of the out-of-span residual $r_t$ (the constant factor drops out of changes). If $L$ has rank $r-1$ and $\Delta\varepsilon^c_j=0$, then $\tilde\mu^{-}_j=(L')^{+}(c_j-\Delta\delta^c-R_c'\Gamma s_j)$, so $\E[\tilde\mu^{-}_j\mid s_j,c_j]=\alpha_0+\Pi c_j+Ds_j$ with $D=-(L')^{+}R_c'\Gamma$. Repeating the proof of part (a), the coefficient on $s_j$ becomes
\[
\phi_t=\Gamma'(r_t-r_b)+D'(\lambda_t-\lambda_b).
\]
The additional term is proportional to $R_c$, the pre-treatment movement of the out-of-span residual over the sub-periods, whose size is of order $S_r^{1/2}$. It vanishes when the shock is exactly in span before treatment ($S_r=0$) and is small relative to the main term when the shock is silent, that is, when $S_r$ is small and $|\ell_t|$ is large after treatment.
\end{remark}

\paragraph{Proof of Proposition~\ref{M-prop:ci}.} Write $\bar\tau-\tau_{\mathcal T}=\mathsf{b}+\sigma Z$ with $|\mathsf{b}|\le B$ and $Z\sim N(0,1)$. Coverage is $\Pr(|Z+\mathsf{b}/\sigma|\le\mathrm{cv}_\alpha(B/\sigma))$. The distribution of $|Z+\rho|$ (the square root of a noncentral $\chi^2_1$ with noncentrality $\rho^2$) is stochastically increasing in $|\rho|$, so the coverage probability is minimised at $|\mathsf{b}|=B$, where it equals $1-\alpha$ by definition of $\mathrm{cv}_\alpha$. Stochastic monotonicity also implies that $\mathrm{cv}_\alpha(\rho)$ is increasing in $\rho$, so the half-length $\sigma\,\mathrm{cv}_\alpha(M\kappa_{\mathcal T}\RMSPE/\sigma)$ is continuous and increasing in $M$. The set of $M$ for which $0\notin\mathrm{CI}$ is therefore an interval $[0,M^*)$, and $M^*$ solves the stated equation by continuity. \qed

\section{Implementation}\label{app:implementation}

\paragraph{Latent factors.} Let $\tilde Y$ be the $T\times J$ donor outcome matrix demeaned by unit (over all periods) and then by period. With $\tilde s=(s-\bar s)/\mathrm{sd}(s)$, each row is replaced by $\tilde Y_{t\cdot}-(\tilde Y_{t\cdot}\tilde s/\tilde s'\tilde s)\tilde s'$, which removes the period-$t$ cross-sectional projection on exposure. With singular value decomposition $UDV'$ of the result, $\hat\lambda_t$ collects a constant and the first $k=2$ columns of $UD/\sqrt T$. The diagnostic $\hat\ell_t$ is the residual of $x_t$ on $(1,\hat\lambda_t')$ fitted over $t\le T_0$, divided by the root mean square of the pre-treatment residuals; $\hat\kappa_t^2=(\hat\kappa^\lambda_t)^2+\hat\ell_t^2$ holds exactly by Corollary~\ref{M-cor:fwl}(a). Phase leverages use phase-averaged factor vectors. Results with $k=1$ and $k=3$ are in Table~\ref{M-tab:robust} of the paper.

\paragraph{Frontier and EBSC.} The programme \eqref{M-eq:frontier} is a convex quadratic programme with linear constraints (for scalar exposure, $|\EI(w)|\le d$ is two linear inequalities). It is solved with \texttt{cvxpy} and the CLARABEL interior-point solver on a grid of 41 tolerances (21 in the simulations) equally spaced between $d_{\min}$ and the synthetic-control imbalance. The unconstrained synthetic-control weights are computed by non-negative least squares with the adding-up constraint imposed through a heavily weighted extra row, and they coincide with the \texttt{cvxpy} solution to $10^{-7}$.

\paragraph{Exposure regression.} $\hat\phi_t$ is the least-squares coefficient on $s_j$ in the cross-donor regression of $Y_{jt}-Y_{j,2011}$ on a constant, $s_j$ and $c_j=(Y_{j,2000}-Y_{j,1996},\,Y_{j,2005}-Y_{j,2001},\,Y_{j,2010}-Y_{j,2006})$, the donor's own growth within each third of 1996--2010. The controls end before the base year 2011, so that their noise is not shared with the dependent variable. Its standard error is the delete-one-donor jackknife, $\{(J-1)J^{-1}\sum_j(\hat\phi_{t(-j)}-\bar\phi_{t(\cdot)})^2\}^{1/2}$. Phase averages and their jackknife errors are computed from the same replicates.

\paragraph{Tuning of EASC$^*$.} The criterion \eqref{M-eq:tuning} is evaluated at each frontier point, with $\hat\kappa^\lambda_t$ from the exposure-orthogonal factors and $\widehat{\mathrm{se}}(\hat\phi_t)$ from the jackknife. In the application the minimum is at the synthetic-control end of the frontier in both pools.

\paragraph{Placebos and noise scale.} For each donor $u$, the full procedure (weights, factors, exposure regression) is repeated with $u$ as the treated unit and the other donors as controls. Iran is never a control. With $\bar\tau_u$ the phase average and $\RMSPE_u$ the pre-treatment fit of placebo $u$, $\hat\sigma=\mathrm{sd}_u(\bar\tau_u/\RMSPE_u)\times\RMSPE_1$; the robust version uses $1.4826\,\mathrm{MAD}_u$. For EASC, $(|\EI(\hat w)|\,\overline{\mathrm{se}}(\hat\phi_{\mathcal T}))^2$ is added to $\hat\sigma^2$. The breakdown value $M^*$ is found by bisection on $[0,5]$.

\paragraph{Standard estimators.} The estimator suite follows \citet{abadie2010} (outcome-only weights), \citet{abadielhour2021}, \citet{benmichael2021} (ridge augmentation), \citet{arkhangelsky2021} (Algorithm 1), \citet{athey2021} (nuclear-norm matrix completion with two-way fixed effects, soft-impute), \citet{xu2017} (interactive fixed effects), \citet{amjad2018} (robust SC) and \citet{fermanpinto2021} (de-meaned SC). Every regularisation parameter is chosen by one rule fixed before estimation: fit on 1996--2007 and minimise the mean squared error of the forecast for Iran over 2008--11. In the simulations, the matrix-completion penalty is set to 1\% of the leading singular value of the outcome matrix, the value the rule selects in the application, to save computation.

\section{Additional Monte Carlo results}\label{app:mc}

The data-generating process is described in Section~\ref{M-sec:mc-design} of the paper. Table~\ref{tab:mc-bias-app} reports bias and RMSE for the 2012--15 and 2016--17 averages, Table~\ref{tab:mc-cover-app} coverage for 2012--15, Table~\ref{tab:mc-cover-mad} coverage with the robust noise scale, and Table~\ref{tab:mc-diag} the diagnostics and the frontier point selected by EASC$^*$.

\begin{table}[htbp]\centering\small
\caption{Monte Carlo: bias and RMSE, 2012--15 and 2016--17}\label{tab:mc-bias-app}
\setlength{\tabcolsep}{4pt}\begin{adjustbox}{max width=\textwidth}\begin{tabular}{@{}lcccccccc@{}}\toprule
 & S0 & S1 & S2 & S3 & S4 & S5 & S6 & S7\\
 & \scriptsize No exposure & \scriptsize In-span shock & \scriptsize Outside hull & \scriptsize Inside hull & \scriptsize Noisy exposure & \scriptsize Concave, outside & \scriptsize Concave, inside & \scriptsize 38 donors\\\midrule
\multicolumn{9}{@{}l}{\emph{Bias, 2012--15 average}}\\
\quad SC & 0.001 & 0.002 & $-$0.046 & $-$0.010 & $-$0.044 & $-$0.026 & $-$0.014 & $-$0.035\\
\quad Augmented SC & 0.000 & 0.002 & $-$0.024 & $-$0.005 & $-$0.023 & $-$0.019 & $-$0.010 & $-$0.017\\
\quad Synthetic DiD & 0.001 & $-$0.001 & $-$0.057 & $-$0.023 & $-$0.055 & $-$0.038 & $-$0.026 & $-$0.048\\
\quad De-meaned SC & 0.000 & 0.001 & $-$0.041 & $-$0.007 & $-$0.039 & $-$0.023 & $-$0.012 & $-$0.032\\
\quad Generalized SC & 0.000 & 0.003 & $-$0.016 & $-$0.003 & $-$0.012 & $-$0.017 & $-$0.011 & $-$0.008\\
\quad Matrix completion & 0.001 & 0.001 & $-$0.068 & $-$0.047 & $-$0.066 & $-$0.038 & $-$0.032 & $-$0.063\\
\quad EBSC-min & $-$0.005 & $-$0.004 & $-$0.021 & $-$0.002 & $-$0.017 & 0.007 & $-$0.006 & $-$0.020\\
\quad EASC & 0.001 & 0.008 & 0.003 & 0.001 & 0.002 & 0.015 & $-$0.005 & 0.007\\
\quad EASC$^*$ & 0.001 & 0.008 & 0.003 & 0.000 & 0.001 & 0.015 & $-$0.005 & 0.007\\
\multicolumn{9}{@{}l}{\emph{RMSE, 2012--15 average}}\\
\quad SC & 0.018 & 0.023 & 0.051 & 0.023 & 0.050 & 0.034 & 0.025 & 0.041\\
\quad Augmented SC & 0.019 & 0.025 & 0.035 & 0.022 & \textbf{0.037} & \textbf{0.029} & 0.023 & 0.030\\
\quad Synthetic DiD & 0.019 & 0.024 & 0.062 & 0.032 & 0.061 & 0.043 & 0.034 & 0.054\\
\quad De-meaned SC & 0.018 & 0.024 & 0.047 & 0.022 & 0.046 & 0.032 & 0.024 & 0.038\\
\quad Generalized SC & 0.020 & 0.030 & 0.041 & 0.028 & 0.040 & 0.035 & 0.030 & 0.035\\
\quad Matrix completion & \textbf{0.017} & \textbf{0.018} & 0.070 & 0.051 & 0.069 & 0.042 & 0.036 & 0.065\\
\quad EBSC-min & 0.316 & 0.333 & 0.331 & 0.030 & 0.301 & 0.330 & 0.032 & 0.159\\
\quad EASC & 0.031 & 0.035 & \textbf{0.033} & 0.021 & 0.048 & 0.035 & 0.021 & 0.025\\
\quad EASC$^*$ & 0.029 & 0.035 & 0.033 & \textbf{0.021} & 0.048 & 0.034 & \textbf{0.021} & \textbf{0.025}\\
\multicolumn{9}{@{}l}{\emph{Bias, 2016--17 average}}\\
\quad SC & 0.000 & 0.001 & $-$0.138 & $-$0.029 & $-$0.134 & $-$0.078 & $-$0.044 & $-$0.108\\
\quad Augmented SC & $-$0.001 & 0.001 & $-$0.071 & $-$0.014 & $-$0.071 & $-$0.056 & $-$0.031 & $-$0.055\\
\quad Synthetic DiD & 0.000 & $-$0.001 & $-$0.181 & $-$0.095 & $-$0.180 & $-$0.114 & $-$0.087 & $-$0.156\\
\quad De-meaned SC & $-$0.001 & 0.001 & $-$0.121 & $-$0.020 & $-$0.119 & $-$0.069 & $-$0.037 & $-$0.099\\
\quad Generalized SC & $-$0.001 & 0.003 & $-$0.049 & $-$0.015 & $-$0.040 & $-$0.054 & $-$0.037 & $-$0.027\\
\quad Matrix completion & 0.000 & 0.001 & $-$0.189 & $-$0.127 & $-$0.187 & $-$0.108 & $-$0.089 & $-$0.178\\
\quad EBSC-min & $-$0.002 & $-$0.007 & $-$0.072 & $-$0.002 & $-$0.082 & $-$0.011 & $-$0.022 & $-$0.069\\
\quad EASC & $-$0.001 & 0.003 & 0.001 & 0.001 & $-$0.007 & 0.036 & $-$0.017 & 0.005\\
\quad EASC$^*$ & $-$0.001 & 0.003 & 0.001 & 0.001 & $-$0.010 & 0.035 & $-$0.018 & 0.005\\
\multicolumn{9}{@{}l}{\emph{RMSE, 2016--17 average}}\\
\quad SC & 0.026 & 0.034 & 0.143 & 0.047 & 0.140 & 0.085 & 0.055 & 0.114\\
\quad Augmented SC & 0.027 & 0.036 & 0.089 & 0.041 & \textbf{0.089} & 0.069 & 0.048 & 0.074\\
\quad Synthetic DiD & 0.025 & 0.031 & 0.184 & 0.100 & 0.184 & 0.117 & 0.091 & 0.159\\
\quad De-meaned SC & 0.026 & 0.035 & 0.128 & 0.042 & 0.125 & 0.077 & 0.050 & 0.105\\
\quad Generalized SC & 0.029 & 0.046 & 0.107 & 0.064 & 0.096 & 0.088 & 0.071 & 0.090\\
\quad Matrix completion & \textbf{0.024} & \textbf{0.024} & 0.190 & 0.130 & 0.189 & 0.111 & 0.092 & 0.179\\
\quad EBSC-min & 0.332 & 0.345 & 0.351 & 0.040 & 0.320 & 0.341 & 0.048 & 0.188\\
\quad EASC & 0.046 & 0.049 & \textbf{0.050} & 0.029 & 0.105 & 0.061 & 0.035 & 0.037\\
\quad EASC$^*$ & 0.041 & 0.049 & 0.050 & \textbf{0.029} & 0.106 & \textbf{0.060} & \textbf{0.035} & \textbf{0.037}\\
\midrule
\quad Mean $\kappa$ (SC), 2018--24 & 9.8 & 6.0 & 9.9 & 10.1 & 9.8 & 10.1 & 9.8 & 10.0\\
\bottomrule\end{tabular}\end{adjustbox}
\par\smallskip\begin{minipage}{0.97\textwidth}\footnotesize\emph{Notes.} 500 replications per design. Errors are phase averages of $\hat\tau_t-\tau_t$; the true effect is $-0.10$ in 2012--15 and $-0.05$ afterwards. Bold: smallest RMSE in the column. Designs: S0 no exposure channel ($\gamma=0$); S1 post-2012 oil price on its pre-2012 relation with the latent trend; S2 actual Brent, treated exposure 22.87 against donor exposures of the 19 normalisers (maximum 16.56); S3 as S2 with 21 donors (adds 26.52 and 38.42); S4 as S2 with exposure observed with multiplicative log-normal error (s.d.\ 0.3); S5 as S2 with concave exposure $\theta_j=\gamma c(1-e^{-s_j/c})$, $c=20$; S6 as S5 with 21 donors; S7 as S2 with 38 donors; S8--S11 are the adverse designs of Section~\ref{M-sec:mc-design}.\end{minipage}
\end{table}

\begin{table}[htbp]\centering\small
\caption{Monte Carlo: coverage and length, 2012--15}\label{tab:mc-cover-app}
\setlength{\tabcolsep}{4pt}\begin{adjustbox}{max width=\textwidth}\begin{tabular}{@{}llcccccccc@{}}\toprule
 & & S0 & S1 & S2 & S3 & S4 & S5 & S6 & S7\\\midrule
\multicolumn{10}{@{}l}{\emph{Coverage of 95\% bias-aware intervals, 2012--15}}\\
\quad SC & $M=0$ & 0.87 & 1.00 & 0.94 & 0.93 & 0.93 & 0.88 & 0.90 & 0.95\\
\quad  & $M=0.5$ & 0.94 & 1.00 & 0.99 & 0.97 & 0.99 & 0.96 & 0.97 & 0.99\\
\quad  & $M=1$ & 0.99 & 1.00 & 1.00 & 1.00 & 1.00 & 0.99 & 0.99 & 1.00\\
\quad EASC & $M=0$ & 0.97 & 0.98 & 1.00 & 0.93 & 0.97 & 0.95 & 0.95 & 0.99\\
\quad  & $M=0.5$ & 0.98 & 1.00 & 1.00 & 0.97 & 0.98 & 0.97 & 0.97 & 1.00\\
\quad  & $M=1$ & 0.99 & 1.00 & 1.00 & 0.99 & 0.99 & 0.99 & 0.99 & 1.00\\
\multicolumn{10}{@{}l}{\emph{Average length, 2012--15}}\\
\quad SC & $M=0$ & 0.066 & 0.161 & 0.183 & 0.093 & 0.182 & 0.116 & 0.093 & 0.166\\
\quad  & $M=0.5$ & 0.090 & 0.207 & 0.244 & 0.123 & 0.241 & 0.155 & 0.124 & 0.213\\
\quad  & $M=1$ & 0.124 & 0.278 & 0.334 & 0.168 & 0.330 & 0.213 & 0.170 & 0.286\\
\quad EASC & $M=0$ & 0.139 & 0.190 & 0.197 & 0.086 & 0.215 & 0.148 & 0.093 & 0.160\\
\quad  & $M=0.5$ & 0.150 & 0.231 & 0.240 & 0.109 & 0.257 & 0.173 & 0.116 & 0.195\\
\quad  & $M=1$ & 0.173 & 0.301 & 0.313 & 0.146 & 0.328 & 0.218 & 0.153 & 0.253\\
\bottomrule\end{tabular}\end{adjustbox}
\par\smallskip\begin{minipage}{0.92\textwidth}\footnotesize\emph{Notes.} Intervals of Proposition~\ref{M-prop:ci} with bias bound $M\hat\kappa\,\RMSPE$ and noise s.d.\ from standardised in-space placebos (Section~\ref{M-sec:inference}); for EASC the jackknife variance of the correction is added. Nominal coverage 0.95. Designs as in Table~\ref{M-tab:mc-bias}.\end{minipage}
\end{table}

\begin{table}[htbp]\centering\small
\caption{Monte Carlo: coverage and length in the adverse designs, 2018--24}\label{tab:mc-cover-adverse}
\setlength{\tabcolsep}{4pt}\begin{adjustbox}{max width=\textwidth}\begin{tabular}{@{}llccccc@{}}\toprule
 & & S3 & S8 & S9 & S10 & S11\\\midrule
\multicolumn{7}{@{}l}{\emph{Coverage of 95\% bias-aware intervals, 2018--24}}\\
\quad SC & $M=0$ & 0.91 & 0.94 & 0.94 & 0.98 & 0.75\\
\quad  & $M=0.5$ & 0.98 & 0.99 & 0.98 & 1.00 & 0.90\\
\quad  & $M=1$ & 1.00 & 1.00 & 1.00 & 1.00 & 0.98\\
\quad EASC & $M=0$ & 0.95 & 0.95 & 0.97 & 0.97 & 0.80\\
\quad  & $M=0.5$ & 0.98 & 0.97 & 0.98 & 1.00 & 0.90\\
\quad  & $M=1$ & 1.00 & 0.99 & 1.00 & 1.00 & 0.97\\
\multicolumn{7}{@{}l}{\emph{Average length, 2018--24}}\\
\quad SC & $M=0$ & 0.213 & 0.211 & 0.598 & 0.980 & 0.253\\
\quad  & $M=0.5$ & 0.296 & 0.296 & 0.823 & 1.346 & 0.334\\
\quad  & $M=1$ & 0.414 & 0.415 & 1.142 & 1.868 & 0.456\\
\quad EASC & $M=0$ & 0.160 & 0.158 & 0.629 & 1.004 & 0.181\\
\quad  & $M=0.5$ & 0.205 & 0.201 & 0.813 & 1.304 & 0.225\\
\quad  & $M=1$ & 0.274 & 0.269 & 1.093 & 1.760 & 0.296\\
\bottomrule\end{tabular}\end{adjustbox}
\par\smallskip\begin{minipage}{0.92\textwidth}\footnotesize\emph{Notes.} Intervals of Proposition~\ref{M-prop:ci} with bias bound $M\hat\kappa\,\RMSPE$ and noise s.d.\ from standardised in-space placebos (Section~\ref{M-sec:inference}); for EASC the jackknife variance of the correction is added. Nominal coverage 0.95. Designs as in Table~\ref{M-tab:mc-bias}.\end{minipage}
\end{table}

\begin{table}[htbp]\centering\small
\caption{Monte Carlo: coverage with the robust (MAD) noise scale}\label{tab:mc-cover-mad}
\setlength{\tabcolsep}{4pt}\begin{adjustbox}{max width=\textwidth}\begin{tabular}{@{}llcccccccc@{}}\toprule
 & & S0 & S1 & S2 & S3 & S4 & S5 & S6 & S7\\\midrule
\multicolumn{10}{@{}l}{\emph{Coverage of 95\% bias-aware intervals, 2016--17}}\\
\quad SC & $M=0$ & 0.84 & 0.99 & 0.58 & 0.87 & 0.59 & 0.60 & 0.79 & 0.72\\
\quad  & $M=0.5$ & 0.97 & 1.00 & 0.92 & 0.97 & 0.94 & 0.89 & 0.94 & 0.96\\
\quad  & $M=1$ & 1.00 & 1.00 & 1.00 & 1.00 & 1.00 & 0.99 & 0.99 & 1.00\\
\quad EASC & $M=0$ & 0.94 & 0.99 & 0.99 & 0.90 & 0.83 & 0.90 & 0.88 & 0.99\\
\quad  & $M=0.5$ & 0.96 & 1.00 & 1.00 & 0.96 & 0.89 & 0.94 & 0.95 & 1.00\\
\quad  & $M=1$ & 0.98 & 1.00 & 1.00 & 0.99 & 0.96 & 0.97 & 0.98 & 1.00\\
\multicolumn{10}{@{}l}{\emph{Average length, 2016--17}}\\
\quad SC & $M=0$ & 0.090 & 0.225 & 0.307 & 0.164 & 0.304 & 0.188 & 0.159 & 0.277\\
\quad  & $M=0.5$ & 0.147 & 0.290 & 0.446 & 0.234 & 0.442 & 0.280 & 0.228 & 0.388\\
\quad  & $M=1$ & 0.219 & 0.390 & 0.635 & 0.330 & 0.628 & 0.401 & 0.322 & 0.544\\
\quad EASC & $M=0$ & 0.195 & 0.268 & 0.275 & 0.124 & 0.333 & 0.210 & 0.136 & 0.218\\
\quad  & $M=0.5$ & 0.210 & 0.325 & 0.333 & 0.156 & 0.388 & 0.244 & 0.167 & 0.267\\
\quad  & $M=1$ & 0.241 & 0.421 & 0.430 & 0.206 & 0.486 & 0.305 & 0.217 & 0.349\\
\multicolumn{10}{@{}l}{\emph{Coverage of 95\% bias-aware intervals, 2018--24}}\\
\quad SC & $M=0$ & 0.89 & 1.00 & 0.61 & 0.90 & 0.64 & 0.64 & 0.83 & 0.74\\
\quad  & $M=0.5$ & 0.98 & 1.00 & 0.95 & 0.97 & 0.94 & 0.90 & 0.96 & 0.96\\
\quad  & $M=1$ & 1.00 & 1.00 & 1.00 & 1.00 & 1.00 & 0.98 & 1.00 & 1.00\\
\quad EASC & $M=0$ & 0.97 & 0.99 & 0.99 & 0.94 & 0.86 & 0.93 & 0.93 & 0.99\\
\quad  & $M=0.5$ & 0.98 & 1.00 & 1.00 & 0.98 & 0.93 & 0.95 & 0.97 & 1.00\\
\quad  & $M=1$ & 0.99 & 1.00 & 1.00 & 1.00 & 0.97 & 0.98 & 0.99 & 1.00\\
\multicolumn{10}{@{}l}{\emph{Average length, 2018--24}}\\
\quad SC & $M=0$ & 0.113 & 0.284 & 0.372 & 0.204 & 0.366 & 0.235 & 0.197 & 0.332\\
\quad  & $M=0.5$ & 0.181 & 0.376 & 0.540 & 0.289 & 0.532 & 0.348 & 0.281 & 0.468\\
\quad  & $M=1$ & 0.267 & 0.512 & 0.767 & 0.406 & 0.757 & 0.498 & 0.396 & 0.658\\
\quad EASC & $M=0$ & 0.258 & 0.342 & 0.351 & 0.159 & 0.410 & 0.281 & 0.172 & 0.275\\
\quad  & $M=0.5$ & 0.277 & 0.422 & 0.431 & 0.204 & 0.487 & 0.329 & 0.216 & 0.344\\
\quad  & $M=1$ & 0.318 & 0.553 & 0.562 & 0.274 & 0.619 & 0.414 & 0.285 & 0.455\\
\bottomrule\end{tabular}\end{adjustbox}
\par\smallskip\begin{minipage}{0.92\textwidth}\footnotesize\emph{Notes.} Intervals of Proposition~\ref{M-prop:ci} with bias bound $M\hat\kappa\,\RMSPE$ and noise s.d.\ from standardised in-space placebos (Section~\ref{M-sec:inference}); for EASC the jackknife variance of the correction is added. Nominal coverage 0.95. Designs as in Table~\ref{M-tab:mc-bias}.\end{minipage}
\end{table}

\begin{table}[htbp]\centering\small
\caption{Monte Carlo: diagnostics and selected frontier points (means over replications)}\label{tab:mc-diag}
\setlength{\tabcolsep}{4pt}\begin{adjustbox}{max width=\textwidth}\begin{tabular}{@{}lcccccccccccc@{}}\toprule
 & S0 & S1 & S2 & S3 & S4 & S5 & S6 & S7 & S8 & S9 & S10 & S11\\\midrule
Pre-treatment $R^2$ of $x_t$ on $\hat\lambda_t$ & 0.890 & 0.892 & 0.891 & 0.894 & 0.887 & 0.891 & 0.890 & 0.895 & 0.893 & 0.893 & 0.892 & 0.892\\
$\max_{t>T_0}|\hat\ell_t|$ & 9.73 & 1.94 & 9.66 & 9.72 & 9.46 & 9.76 & 9.43 & 9.79 & 9.87 & 15.22 & 14.94 & 9.87\\
$d_{\min}$ & 6.31 & 6.31 & 6.31 & 0.00 & 6.11 & 6.31 & 0.00 & 6.31 & 0.00 & 0.00 & 0.00 & 0.00\\
EI of SC & 19.87 & 11.65 & 11.70 & 2.46 & 12.01 & 13.51 & 4.65 & 9.40 & 2.38 & 2.33 & 3.00 & 2.39\\
EI of EASC$^*$ & 15.77 & 11.42 & 11.51 & 2.34 & 11.51 & 12.68 & 3.91 & 9.33 & 2.27 & 2.19 & 2.94 & 2.27\\
EASC$^*$: frontier position (1 = SC) & 0.70 & 0.96 & 0.96 & 0.95 & 0.92 & 0.89 & 0.85 & 0.97 & 0.95 & 0.94 & 0.96 & 0.94\\
RMSPE of SC & 0.0087 & 0.0226 & 0.0231 & 0.0116 & 0.0229 & 0.0149 & 0.0117 & 0.0189 & 0.0119 & 0.0119 & 0.0205 & 0.0119\\
RMSPE of EBSC-min & 0.2512 & 0.2682 & 0.2635 & 0.0167 & 0.2146 & 0.2601 & 0.0166 & 0.1090 & 0.0164 & 0.0162 & 0.0284 & 0.0167\\
\bottomrule\end{tabular}\end{adjustbox}
\end{table}


\section{Treatment, donors and data}\label{app:donors}

\subsection{Treatment timing}

\begin{table}[H]\centering\small
\caption{Dated sequence of the confrontation}\label{tab:timeline}
\begin{tabularx}{\textwidth}{@{}lLL@{}}\toprule
Date & Measure or event & Role in the design\\\midrule
Aug 2005 & New executive takes office & Inside the pre-period\\
Dec 2006 & UN Security Council Resolution 1737 & Start of earlier designs (treatment from 2007)\\
Jun 2009 & Disputed election and crackdown & Inside the pre-period\\
Jun--Jul 2010 & UNSC Resolution 1929; US CISADA & Backdated treatment (from 2010)\\
Dec 2010 & Targeted subsidy reform & Domestic price shock before treatment\\
Nov--Dec 2011 & US Section 311 finding; NDAA FY2012, Sec.\ 1245 (central bank) & Announcement; backdated treatment (from 2011)\\
Jan 2012 & EU oil embargo decided (Council Decision 2012/35/CFSP) & \textbf{Treatment}\\
Mar 2012 & SWIFT disconnects designated banks; EU Regulation 267/2012 & \textbf{Treatment}\\
Jul 2012 & EU oil embargo in force & \textbf{Treatment}\\
Nov 2013 & Joint Plan of Action & Negotiations open\\
Jul 2015 / Jan 2016 & JCPOA agreed / implemented & Relief phase 2016--17\\
May--Nov 2018 & US withdrawal; sanctions reimposed & Siege phase begins\\
May 2019 & End of US waivers for buyers of Iranian oil & Re-intensification\\
\bottomrule\end{tabularx}
\end{table}

\subsection{Donor-pool criteria}

A country enters the baseline (normaliser) pool only if it meets all six criteria:
\begin{enumerate}[itemsep=0pt]
\item formal diplomatic relations with the United States throughout 1996--2011;
\item formal diplomatic relations with Israel throughout 1996--2011, the criterion that separates accommodation from confrontation in Iran's regional setting;
\item an institutional framework with the European Union in force by 2011 (membership; an association, partnership or cooperation agreement; or the Cotonou framework);
\item no comprehensive, trade or financial sanctions by the UN, US or EU in 1996--2024 (arms embargoes, aid suspensions and individual designations are recorded but do not disqualify);
\item no successful coup, revolution or new civil war in 2010--13;
\item no land border with Iran and no documented role as a sanctions-evasion hub.
\end{enumerate}
Thirteen countries were specified first; six Latin American countries were added to raise the number of placebo units. The exposure-feasible pool adds Azerbaijan and Angola, which meet criteria 1--5. Azerbaijan fails criterion 6 (it borders Iran); Angola was excluded from the baseline because of its civil war, which ended in 2002. In the exposure-feasible pool both receive zero synthetic-control weight. They enter the exposure regression, the latent-factor estimates and the placebo distribution, and Angola receives the largest weight (0.49) in EBSC-min, the exposure-balanced weights. Table~\ref{M-tab:robust} in the paper shows that the results are the same with Angola alone.

\begin{table}[htbp]\centering\scriptsize
\caption{Donor pool and the evidence for each decision}\label{tab:donors}
\begin{tabularx}{\textwidth}{@{}l>{\raggedright\arraybackslash}p{1.0cm}>{\raggedright\arraybackslash}p{2.2cm}>{\raggedright\arraybackslash}p{2.0cm}>{\raggedright\arraybackslash}p{2.6cm}L>{\raggedright\arraybackslash}p{1.6cm}@{}}\toprule
Country & Tier & United States & Israel & EU framework in force & Flags & Decision\\\midrule
Kazakhstan & Oil & Since 1991--92 & Since 1992 & PCA, 1999 & Caspian neighbour; oil swaps through Iran in the 2000s & Normaliser\\
Nigeria & Oil & Continuous & Restored 1992 & Cotonou, 2000 & EU arms embargo and visa bans 1995--98; insurgency from 2009 & Normaliser\\
Ecuador & Oil & Continuous; ambassadors expelled 2011, restored 2012 & Continuous & EC--Andean framework, 1998 & 1999--2000 banking crisis & Normaliser\\
Mexico & Oil & Continuous & Since 1952 & Global Agreement, 2000 & -- & Normaliser\\
Colombia & Oil & Continuous & Continuous to May 2024 & EC--Andean framework, 1998 & US decertification 1996--97; armed conflict to 2016 & Normaliser\\
Brazil & Oil & Continuous & Since 1949 & Framework agreement, 1995 & Co-sponsored 2010 Tehran fuel-swap declaration & Normaliser (ext.)\\
Jordan & Non-oil & Continuous; FTA 2001 & Peace treaty, Oct 1994 & Association Agreement, 2002 & Syrian refugee inflow after 2012 & Normaliser\\
Romania & Non-oil & Continuous & Continuous since 1948 & Europe Agreement 1995; EU 2007 & EU convergence & Normaliser\\
Bulgaria & Non-oil & Continuous & Restored 1990 & Europe Agreement 1995; EU 2007 & 1996--97 crisis & Normaliser\\
Vietnam & Non-oil & Normalised Jul 1995 & Since 1993 & Cooperation agreement, 1996 & Transition economy & Normaliser\\
South Africa & Non-oil & Continuous & Continuous; ambassador recalled 2018 & TDCA, 2000/2004 & -- & Normaliser\\
Philippines & Non-oil & Treaty ally & Since 1957 & EC--ASEAN, 1980 & -- & Normaliser\\
Thailand & Non-oil & Treaty ally & Since 1954 & EC--ASEAN, 1980 & Coups 2006 and 2014 & Normaliser\\
Kenya & Non-oil & Continuous & Restored 1988 & Cotonou, 2000 & -- & Normaliser\\
Chile & Non-oil & Continuous & Continuous & Association Agreement, 2003 & -- & Normaliser (ext.)\\
Peru & Non-oil & Continuous & Continuous & EC--Andean framework, 1998 & -- & Normaliser (ext.)\\
Costa Rica & Non-oil & Continuous & Continuous & EC--Central America framework, 1999 & -- & Normaliser (ext.)\\
Dominican Rep. & Non-oil & Continuous & Continuous & Cotonou, 2000; EPA 2008 & -- & Normaliser (ext.)\\
Panama & Non-oil & Continuous & Continuous & EC--Central America framework, 1999 & -- & Normaliser (ext.)\\\midrule
Azerbaijan & Oil & Since 1992 & Since 1992 & PCA, 1999 & Borders Iran & Exposure support only\\
Angola & Oil & Since 1993 & Since 1992 & Cotonou, 2000 & Civil war to 2002 & Exposure support only\\
\bottomrule\end{tabularx}
\par\smallskip\footnotesize\emph{Notes.} ``(ext.)'': Latin American extension added after the core thirteen were specified. Excluded candidates and reasons: Indonesia, Malaysia, Morocco and Tunisia (no relations with Israel over 1996--2011); Egypt (2011 revolution, 2013 coup); Turkey (gas-for-gold trade with Iran 2012--13); Serbia (sanctions 1998--2000); Iraq, Afghanistan, Armenia and Pakistan (border); China, India and the United Arab Emirates (buyers of Iranian crude or evasion hubs).
\end{table}

\begin{table}[htbp]\centering\small
\caption{Donors: oil exposure and weights}\label{tab:weights}
\begin{adjustbox}{max width=\textwidth}\begin{tabular}{@{}lcccc@{}}\toprule
Donor & Oil rents, \% GDP & SC weight & EBSC-min weight & EBSC-min weight\\
 & 1996--2011 & & (normalisers) & (exposure-feasible)\\\midrule
Angola & 38.42 & -- & -- & 0.488\\
Azerbaijan & 26.52 & -- & -- & 0.000\\
Kazakhstan & 16.56 & 0.133 & 1.000 & 0.172\\
Nigeria & 13.85 & 0.254 & 0.000 & 0.000\\
Ecuador & 11.56 & 0.000 & 0.000 & 0.000\\
Vietnam & 5.48 & 0.000 & 0.000 & 0.000\\
Colombia & 4.09 & 0.000 & 0.000 & 0.000\\
Mexico & 3.79 & 0.369 & 0.000 & 0.340\\
Brazil & 1.51 & 0.000 & 0.000 & 0.000\\
Peru & 1.08 & 0.000 & 0.000 & 0.000\\
Romania & 1.05 & 0.000 & 0.000 & 0.000\\
Thailand & 0.89 & 0.243 & 0.000 & 0.000\\
South Africa & 0.73 & 0.000 & 0.000 & 0.000\\
Philippines & 0.11 & 0.000 & 0.000 & 0.000\\
Costa Rica & 0.10 & 0.000 & 0.000 & 0.000\\
Chile & 0.04 & 0.000 & 0.000 & 0.000\\
Bulgaria & 0.03 & 0.000 & 0.000 & 0.000\\
Kenya & 0.00 & 0.000 & 0.000 & 0.000\\
Jordan & 0.00 & 0.000 & 0.000 & 0.000\\
Dominican Rep. & 0.00 & 0.000 & 0.000 & 0.000\\
Panama & 0.00 & 0.000 & 0.000 & 0.000\\
\midrule
Iran & 22.87 &  &  & \\
Weighted donor exposure &  & 7.35 & 16.56 & 22.87\\
\bottomrule\end{tabular}\end{adjustbox}
\par\smallskip\begin{minipage}{0.8\textwidth}\footnotesize\emph{Notes.} Oil rents: WDI NY.GDP.PETR.RT.ZS; the Dominican Republic and Panama report no oil rents and are set to zero. SC weights are identical in the two pools because Azerbaijan and Angola receive zero weight.\end{minipage}
\end{table}


\subsection{Data}

\begin{table}[H]\centering\small
\caption{Data sources}\label{tab:data}
\begin{tabularx}{\textwidth}{@{}lLL@{}}\toprule
Variable & Source & Notes\\\midrule
GDP per capita, PPP, constant 2021 \$ & World Development Indicators (NY.GDP.PCAP.PP.KD) & Logs; 1996--2024\\
Oil rents, \% of GDP & WDI (NY.GDP.PETR.RT.ZS) & Mean 1996--2011 (baseline) or 2002--2011; missing values for the Dominican Republic and Panama set to zero\\
Oil price & Brent, US\$ per barrel, annual average (FRED series POILBREUSDA) & Logs\\
Petroleum, coal and gas income & V-Dem v16, distributed series of \citet{haber2011} & Mean 1996--2006; alternative exposure index\\
Iranian crude exports & \citet{eia2015,eia2026} & Context only\\
\bottomrule\end{tabularx}
\end{table}

\noindent The WDI series for several donors were retrieved through the World Bank API from a cloud environment; the replication package includes \texttt{get\_wdi.py}, which re-downloads every series with \texttt{wbgapi}. Official statistics in autocracies may overstate growth \citep{martinez2022}. If Iran's post-2020 national accounts are inflated, the positive EASC gaps after 2020 are overstated.

\section{Additional empirical results}\label{app:results}

\begin{table}[htbp]\centering\small
\caption{Standard estimators before and after exposure augmentation, normaliser pool}\label{tab:estimators-base}
\begin{adjustbox}{max width=\textwidth}\begin{tabular}{@{}lccccccc@{}}\toprule
 & & \multicolumn{3}{c}{As published} & \multicolumn{3}{c}{Exposure-augmented}\\
\cmidrule(lr){3-5}\cmidrule(lr){6-8}
Estimator & EI($\hat w$) & 2012--15 & 2016--17 & 2018--24 & 2012--15 & 2016--17 & 2018--24\\\midrule
SC & 15.5 & $-$0.127 & $-$0.087 & $-$0.089 & $-$0.152 & $-$0.033 & 0.086\\
SC with covariates, oil rents a predictor & 13.3 & $-$0.119 & $-$0.065 & $-$0.052 & $-$0.140 & $-$0.019 & 0.098\\
Penalized SC & 22.0 & $-$0.119 & $-$0.098 & $-$0.118 & $-$0.154 & $-$0.021 & 0.131\\
Augmented SC & 14.4 & $-$0.114 & $-$0.053 & $-$0.024 & $-$0.136 & $-$0.003 & 0.139\\
Synthetic DiD & 18.4 & $-$0.113 & $-$0.076 & $-$0.111 & $-$0.142 & $-$0.012 & 0.097\\
De-meaned SC & 14.8 & $-$0.128 & $-$0.081 & $-$0.073 & $-$0.152 & $-$0.029 & 0.095\\
Matrix completion & -- & $-$0.142 & $-$0.145 & $-$0.192 & -- & -- & --\\
Generalized SC & -- & $-$0.034 & 0.164 & 0.267 & -- & -- & --\\
Robust SC & -- & $-$0.040 & 0.137 & 0.221 & -- & -- & --\\
\bottomrule\end{tabular}\end{adjustbox}
\par\smallskip\begin{minipage}{0.95\textwidth}\footnotesize\emph{Notes.} As Table~\ref{M-tab:estimators}, with the 19 normalisers as donors and $\hat\phi_t$ estimated on the same 19 donors, so that the correction extrapolates beyond the donors' exposure range.\end{minipage}
\end{table}

\begin{table}[htbp]\centering\small
\caption{Leverage and breakdown values by number of latent factors}\label{tab:factors-k}
\begin{adjustbox}{max width=\textwidth}\begin{tabular}{@{}llccccc@{}}\toprule
Pool & $k$ & $R^2$ & $\kappa$ 2012--15 & $\kappa$ 2016--17 & $\kappa$ 2018--24 & $M^*$ (SC, 2012--15)\\\midrule
Normalisers & 1 & 0.920 & 3.4 & 9.7 & 11.5 & 0.96\\
 & 2 & 0.921 & 7.9 & 18.9 & 26.7 & 0.41\\
 & 3 & 0.922 & 10.9 & 34.2 & 56.4 & 0.30\\
Exposure-feasible & 1 & 0.915 & 2.2 & 6.3 & 6.9 & 1.48\\
 & 2 & 0.915 & 12.1 & 28.1 & 40.4 & 0.27\\
 & 3 & 0.918 & 14.5 & 37.5 & 56.8 & 0.23\\
\bottomrule\end{tabular}\end{adjustbox}
\par\smallskip\begin{minipage}{0.9\textwidth}\footnotesize\emph{Notes.} $k$ principal components of two-way demeaned, exposure-orthogonalised donor outcomes; the paper uses $k=2$. $\kappa$ is the synthetic-control leverage of the phase average; $M^*$ is the breakdown value of the 2012--15 synthetic-control estimate.\end{minipage}
\end{table}


\begin{figure}[htbp]\centering
\includegraphics[width=\textwidth]{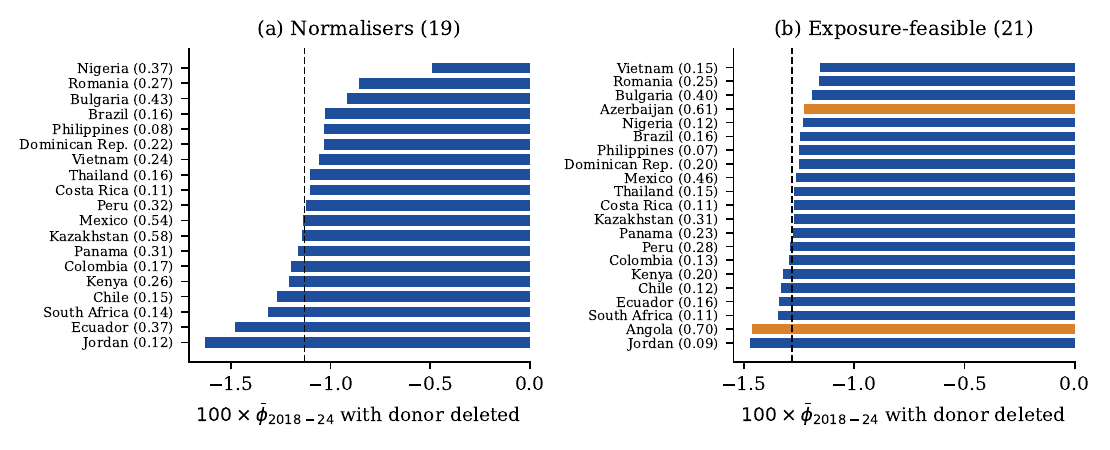}
\caption{Influence of each donor on the exposure regression: 2018--24 average of $\hat\phi_t$ with the donor deleted (bars), full-sample value (dashed line), and the donor's hat value in parentheses. Orange: the two high-exposure donors.}
\label{fig:influence}
\end{figure}
\begin{table}[htbp]\centering\small
\caption{In-time placebos around the 2008--09 oil collapse (sample 1996--2011)}\label{tab:intime}
\begin{adjustbox}{max width=\textwidth}\begin{tabular}{@{}llcccccccc@{}}\toprule
 & Placebo & & & & \multicolumn{2}{c}{Mean gap, placebo period} & \multicolumn{2}{c}{Mean $|$gap$|$}\\
\cmidrule(lr){6-7}\cmidrule(lr){8-9}
Pool & treated year & EI & RMSPE & $\max|\ell_t|$ & SC & EASC & SC & EASC\\\midrule
Normalisers & 2006 & 18.3 & 0.014 & 1.9 & $-$0.069 & $-$0.007 & 0.069 & 0.029\\
Normalisers & 2007 & 15.5 & 0.016 & 2.0 & $-$0.008 & $-$0.005 & 0.009 & 0.028\\
Normalisers & 2008 & 15.7 & 0.016 & 2.2 & $-$0.012 & $-$0.019 & 0.012 & 0.023\\
Normalisers & 2009 & 16.1 & 0.016 & 2.5 & $-$0.003 & $-$0.049 & 0.005 & 0.049\\
Exposure-feasible & 2006 & 18.3 & 0.014 & 8.1 & $-$0.069 & $-$0.099 & 0.069 & 0.099\\
Exposure-feasible & 2007 & 15.5 & 0.016 & 5.0 & $-$0.008 & 0.011 & 0.009 & 0.012\\
Exposure-feasible & 2008 & 15.7 & 0.016 & 3.2 & $-$0.012 & $-$0.005 & 0.012 & 0.006\\
Exposure-feasible & 2009 & 16.1 & 0.016 & 3.5 & $-$0.003 & 0.004 & 0.005 & 0.004\\
\bottomrule\end{tabular}\end{adjustbox}
\par\smallskip\begin{minipage}{0.95\textwidth}\footnotesize\emph{Notes.} Data end in 2011. Exposure is the mean of oil rents up to the year before the placebo date; the base year of $\hat\phi_t$ is the last placebo pre-treatment year. No treatment occurs in the placebo window in the paper's design; earlier UN and US measures (2006--10) are part of the pre-treatment period by assumption, so non-zero placebo gaps reflect estimator error or effects of those measures.\end{minipage}
\end{table}

\begin{table}[htbp]\centering\small
\caption{Backdating the treatment}\label{tab:backdate}
\begin{adjustbox}{max width=\textwidth}\begin{tabular}{@{}llccccccc@{}}\toprule
 & & \multicolumn{3}{c}{SC} & \multicolumn{3}{c}{EASC} & EASC\\
\cmidrule(lr){3-5}\cmidrule(lr){6-8}
Pool & First treated year & 2012--15 & 2016--17 & 2018--24 & 2012--15 & 2016--17 & 2018--24 & to 2011\\\midrule
Normalisers & 2009 & $-$0.129 & $-$0.090 & $-$0.092 & $-$0.274 & $-$0.206 & $-$0.100 & $-$0.049\\
Normalisers & 2010 & $-$0.130 & $-$0.091 & $-$0.093 & $-$0.238 & $-$0.166 & $-$0.074 & $-$0.044\\
Normalisers & 2011 & $-$0.129 & $-$0.090 & $-$0.092 & $-$0.185 & $-$0.062 & 0.061 & $-$0.039\\
Normalisers & 2012 & $-$0.127 & $-$0.087 & $-$0.089 & $-$0.152 & $-$0.033 & 0.086 & --\\
Exposure-feasible & 2009 & $-$0.129 & $-$0.090 & $-$0.092 & $-$0.101 & 0.014 & 0.134 & 0.004\\
Exposure-feasible & 2010 & $-$0.130 & $-$0.091 & $-$0.093 & $-$0.098 & 0.015 & 0.129 & 0.005\\
Exposure-feasible & 2011 & $-$0.129 & $-$0.090 & $-$0.092 & $-$0.108 & 0.005 & 0.126 & 0.004\\
Exposure-feasible & 2012 & $-$0.127 & $-$0.087 & $-$0.089 & $-$0.122 & $-$0.011 & 0.109 & --\\
\bottomrule\end{tabular}\end{adjustbox}
\par\smallskip\begin{minipage}{0.95\textwidth}\footnotesize\emph{Notes.} The first treated year is moved earlier; weights, exposure (mean up to the year before) and $\hat\phi_t$ (base year: the year before) are re-estimated. The last column is the mean EASC gap between the backdated first treated year and 2011, before the actual treatment.\end{minipage}
\end{table}

\begin{table}[htbp]\centering\small
\caption{Leave one donor out}\label{tab:loo}
\begin{adjustbox}{max width=\textwidth}\begin{tabular}{@{}llcccccccc@{}}\toprule
 & & & & \multicolumn{3}{c}{SC} & \multicolumn{3}{c}{EASC}\\
\cmidrule(lr){5-7}\cmidrule(lr){8-10}
Pool & Dropped & Weight & EI & 2012--15 & 2016--17 & 2018--24 & 2012--15 & 2016--17 & 2018--24\\\midrule
Normalisers & Mexico & 0.369 & 20.9 & $-$0.107 & $-$0.055 & $-$0.043 & $-$0.138 & 0.023 & 0.195\\
Normalisers & Nigeria & 0.254 & 19.1 & $-$0.138 & $-$0.137 & $-$0.198 & $-$0.194 & $-$0.129 & $-$0.104\\
Normalisers & Thailand & 0.243 & 14.7 & $-$0.116 & $-$0.071 & $-$0.071 & $-$0.144 & $-$0.027 & 0.092\\
Normalisers & Kazakhstan & 0.133 & 19.8 & $-$0.145 & $-$0.138 & $-$0.190 & $-$0.182 & $-$0.084 & 0.036\\
Exposure-feasible & Mexico & 0.369 & 20.9 & $-$0.107 & $-$0.055 & $-$0.043 & $-$0.101 & 0.047 & 0.221\\
Exposure-feasible & Nigeria & 0.254 & 19.1 & $-$0.138 & $-$0.137 & $-$0.198 & $-$0.133 & $-$0.046 & 0.037\\
Exposure-feasible & Thailand & 0.243 & 14.7 & $-$0.116 & $-$0.071 & $-$0.071 & $-$0.113 & $-$0.001 & 0.116\\
Exposure-feasible & Kazakhstan & 0.133 & 19.8 & $-$0.145 & $-$0.138 & $-$0.190 & $-$0.139 & $-$0.041 & 0.062\\
\bottomrule\end{tabular}\end{adjustbox}
\par\smallskip\begin{minipage}{0.95\textwidth}\footnotesize\emph{Notes.} Each row drops one donor with SC weight above 0.01 and re-estimates weights and $\hat\phi_t$.\end{minipage}
\end{table}


\section{Replication}\label{app:replication}

All computations are in Python 3 (\texttt{numpy}, \texttt{scipy}, \texttt{pandas}, \texttt{cvxpy} with CLARABEL, \texttt{matplotlib}). The package contains:
\begin{itemize}[itemsep=0pt]
\item \texttt{ebsc.py}: diagnostics ($\hat\kappa_t$, $\hat\ell_t$, $\EI$), the exposure--fit frontier (whose first point is EBSC-min), EASC, EASC$^*$, bias-aware intervals and breakdown values;
\item \texttt{modern.py}, \texttt{scm.py}: the standard estimators;
\item \texttt{app.py}, \texttt{app\_inference.py}, \texttt{app\_robust.py}: every number in Section~\ref{M-sec:app};
\item \texttt{mc.py}, \texttt{mc\_summary.py}: the Monte Carlo (seed 20261002, 500 replications per design; the three adverse designs are run by the same script with the \texttt{mc\_extra} tag);
\item \texttt{make\_app\_outputs.py}, \texttt{make\_mc\_outputs.py}: all tables and figures;
\item \texttt{run\_all.sh}: runs everything in order (about 40 minutes on two cores).
\end{itemize}

\bibliographystyle{plainnat}
\bibliography{refs}